\documentclass{emulateapj} 
\usepackage{apjfonts}
\usepackage{epsfig}
\usepackage{lscape}
\usepackage{rotating}

\shorttitle{Starburst Galaxies with Spitzer/IRS}
\shortauthors{Bernard-Salas et al.}

\begin{document}

\title{A Spitzer high resolution mid-infrared spectral atlas of starburst galaxies}

\author{J. Bernard-Salas\altaffilmark{1},
H.W.W. Spoon\altaffilmark{1},
V. Charmandaris\altaffilmark{2,3},
V. Lebouteiller\altaffilmark{1},
D. Farrah\altaffilmark{1,4},
D. Devost\altaffilmark{5},
B.R. Brandl\altaffilmark{6},
Yanling Wu\altaffilmark{1,7}
L. Armus\altaffilmark{7},
L. Hao\altaffilmark{1,8},
G.C. Sloan\altaffilmark{1},
D. Weedman\altaffilmark{1},
J.R. Houck\altaffilmark{1}}

\altaffiltext{1}{Cornell University, 222 Space Sciences Bld., Ithaca,
  NY 14853, USA}

\altaffiltext{2}{University of Crete, Department of Physics, GR-71003,
  Heraklion, Greece}

\altaffiltext{3}{IESL/Foundation for Research \& Technology - Hellas,
  GR-71110, Heraklion, Greece, and Chercheur Associ\'e, Observatoire
  de Paris, F-75014, Paris, France}

\altaffiltext{4}{Department of Physics and Astronomy, University of
  Sussex, Brighton, East Sussex BN1 9QH}

\altaffiltext{5}{CFHT Corporation, 65-1238 Mamalahoa Hwy, Kamuela,
  Hawaii 96743, USA}

\altaffiltext{6}{Leiden Observatory, Leiden University, P.O. Box 9513,
  2300 RA Leiden, The Netherlands}

\altaffiltext{7}{{\em Spitzer} Science Center, MS 220-06, California
  Institute of Technology, Pasadena, CA, 91125}

\altaffiltext{8}{University of Texas at Austin, Department of
  Astronomy, Austin, Texas 78712-0259}

\begin{abstract}
  We present an atlas of {\em Spitzer}/IRS high resolution
  (R$\sim$600) 10-37$\mu$m spectra for 24 well known starburst
  galaxies. The spectra are dominated by fine-structure lines,
  molecular hydrogen lines, and emission bands of polycyclic aromatic
  hydrocarbons.  Six out of the eight objects with a known AGN
  component show emission of the high excitation \ion{[Ne}{5]} line.
  This line is also seen in one other object (NGC\,4194) with, a
  priori, no known AGN component. In addition to strong polycyclic
  aromatic hydrocarbon emission features in this wavelength
  range (11.3, 12.7, 16.4~$\mu$m), the spectra reveal other weak
  hydrocarbon features at 10.6, 13.5, 14.2~$\mu$m, and a previously
  unreported emission feature at 10.75$\mu$m. An unidentified
  absorption feature at 13.7$\mu$m is detected in many of the
  starbursts.  We use the fine-structure lines to derive the abundance
  of neon and sulfur for 14 objects where the HI 7-6 line is
  detected. We further use the molecular hydrogen
  lines to sample the properties of the warm molecular gas.  Several
  basic diagrams characterizing the properties of the sample are also
  shown.  We have combined the spectra of all the pure starburst
  objects to create a high S/N template, which is available to the
  community.
\end{abstract}


\section{Introduction}

As galaxies evolve, they often go through periods when their star
formation rates increase substantially. This is usually ascribed to
large-scale perturbations in their structure, such as the formation of
bars triggered by interactions with companions, leading to the
accumulation of large quantities of gas in their nuclei. A small
number of those so-called `starburst' galaxies
\citep[e.g.][]{Weedman81} have been identified in the local Universe
and have been studied extensively at all wavelengths. At high
redshifts, obscured starburst galaxies are much more
common \citep{Blain02,Elbaz03}, and are believed to give rise to the
bulk of the cosmic infrared background radiation.  Reviews of their
properties can be found in \citet{ken98,lag05} and \citet{lfs06}.

Understanding the physics of starburst galaxies in the local Universe
is of great importance, as they are the closest laboratories we have
for studying the starburst phenomenon, and because they may represent
`scaled-down' versions of the starbursts seen at high redshift.  The
obscured nature of (most) starburst galaxies, regardless of their
redshifts, however means that probing their nuclear regions in 
  the optical is challenging. This task is better done in the
infrared, where dust opacity is lower. The advent of major infrared
space facilities such as {\em IRAS} \citep{neu04}, {\em ISO}
\citep{kess96} and {\em Spitzer} \citep{wer04}, and in particular the
accessibility to high quality spectroscopy, has allowed great progress
in this field, including detailed studies of the mid-infrared (MIR)
region of the spectrum in galaxies. It was discovered that the
  mid-infrared (MIR) is rich in emission and absorption features of
gas and dust. Emission from Polycyclic Aromatic Hydrocarbons
\citep[PAHs,][]{Leger84} was shown to be a reliable star formation
tracer \citep[e.g.][]{Roussel01,Forster04,Peeters04}, though
variations with metallicity have also been observed
\citep{Wu06,Smith07,Rosenberg08}.  The use of infrared atomic lines
with different degrees of ionization has proven invaluable in studying
the properties of the ionization field in starbursts
\citep[e.g.][]{sturmetal02,vermaetal03,Wu06,Dale06}. In ultra-luminous
infrared galaxies (ULIRGs), diagnostic diagrams based on line ratios
such as [NeV$\lambda14.32\mu$m]/[NeII$\lambda12.81\mu$m],
[OIV$\lambda25.89\mu$m]/[NeII$\lambda12.81\mu$m],
[SIII$\lambda33.48\mu$m]/[SiII$\lambda34.82\mu$m] and combinations of
those with PAH emission features have been developed to reveal their
nature
\citep[e.g.][]{genzeletal98,Laurent00,thornleyetal00,Dale06,Armus07,Farrah07,
  Farrah08}. Additional diagnostics using the silicate absorption
features \citep{Spoon07} or the strength of the rotational molecular
hydrogen lines \citep[e.g.][]{rigopoulouetal99,Roussel07} have also
been proposed.

\begin{deluxetable*}{llrrrrrrcc}
\tabletypesize{\scriptsize}
\tablecaption{The Starburst Galaxy Sample \label{sampletable}}

\tablewidth{0pt} \tablehead{ \colhead{Name} & \colhead{Type} &
\colhead{RA} & \colhead{DEC} & \colhead{z} & \colhead{D} &
\colhead{log[L$_{\rm IR}$]} & \colhead{AOR ID\tablenotemark{a}} &
\colhead{SH t$_{int}$} & \colhead{LH t$_{int}$} \\ \colhead{} &
\colhead{} & \colhead{} & \colhead{} & \colhead{} & \colhead{Mpc} &
\colhead{L$_{\odot}$} & \colhead{} & \colhead{(cycles $\times$ s)} &
\colhead{(cycles $\times$ s)} }

\startdata
NGC\,253 & SB     & 00$^h$47$^m$33.12$^s$ & -25$^d$17$^m$17.6$^s$ & 0.0008  &  2.5 & 10.23 & 9072640 & 12 $\times$ 6 & 12 $\times$ 6 \\
NGC\,520 & SB     & 01$^h$24$^m$35.07$^s$ & +03$^d$47$^m$32.7$^s$ & 0.0076  & 30.2 & 10.91 & 9073408 & 4 $\times$ 30 & 3 $\times$ 60 \\
NGC\,660 & SB+AGN\tablenotemark{b} & 01$^h$43$^m$02.35$^s$ & +13$^d$38$^m$44.4$^s$ & 0.00283 & 12.3 & 10.49 & 9070848 & 4 $\times$ 30 & 8 $\times$ 14 \\
NGC\,1097 & SB+AGN\tablenotemark{b} & 02$^h$46$^m$19.08$^s$ & -30$^d$16$^m$28.0$^s$ & 0.00456 & 16.8 & 10.71 & 3758080 & 4 $\times$ 30 & 2 $\times$ 60 \\
NGC\,1222 & SB     & 03$^h$08$^m$56.74$^s$ & -02$^d$57$^m$18.5$^s$ & 0.00818 & 32.3 & 10.60 & 9071872 & 4 $\times$ 30 & 2 $\times$ 60 \\
NGC\,1365 & SB+AGN\tablenotemark{b} & 03$^h$33$^m$36.37$^s$ & -36$^d$08$^m$25.5$^s$ & 0.00545 & 17.9 & 11.00 & 8767232 & 4 $\times$ 30 & 8 $\times$ 14 \\
IC\,342   & SB     & 03$^h$46$^m$48.51$^s$ & +68$^d$05$^m$46.0$^s$ & 0.00010  &  4.6 & 10.17 & 9072128 & 4 $\times$ 30 & 8 $\times$ 14 \\   
NGC\,1614 & SB     & 04$^h$33$^m$59.85$^s$ & -08$^d$34$^m$44.0$^s$ & 0.016   & 62.6 & 11.60 & 3757056 & 4 $\times$ 30 & 8 $\times$ 14 \\
NGC\,2146 & SB     & 06$^h$18$^m$37.71$^s$ & +78$^d$21$^m$25.3$^s$ & 0.003   & 16.5 & 11.07 & 9074432 & 1 $\times$ 6 & 1 $\times$ 60 \\
NGC\,2623 & SB+AGN\tablenotemark{b} & 08$^h$38$^m$24.08$^s$ & +25$^d$45$^m$16.9$^s$ & 0.01846 & 77.4 & 11.54 & 9072896 & 4 $\times$ 30 & 2 $\times$ 60 \\
NGC\,3079 & SB+AGN\tablenotemark{b} & 10$^h$01$^m$57.80$^s$ & +55$^d$40$^m$47.1$^s$ & 0.00372 & 17.6 & 10.52 & 3755520 & 4 $\times$ 30& 2  $\times$ 60\\
NGC\,3256 & SB     & 10$^h$27$^m$51.27$^s$ & -43$^d$54$^m$13.8$^s$ & 0.00913 & 35.4 & 11.56 & 9073920 & 1 $\times$ 6 &  1 $\times$ 60\\
NGC\,3310 & SB     & 10$^h$38$^m$45.96$^s$ & +53$^d$30$^m$05.3$^s$ & 0.0033  & 19.8 & 10.61 & 9071616 & 4 $\times$ 30 & 2 $\times$ 60 \\
NGC\,3556 & SB     & 11$^h$11$^m$30.97$^s$ & +55$^d$40$^m$26.8$^s$ & 0.00233 & 13.9 & 10.37 & 9070592 & 4 $\times$ 30 & 2 $\times$ 60 \\
NGC\,3628 & SB+AGN\tablenotemark{b} & 11$^h$20$^m$17.02$^s$ & +13$^d$35$^m$22.2$^s$ & 0.00281 & 10.0 & 10.25 & 9070080 & 4 $\times$ 30 & 2 $\times$ 60 \\
NGC\,4088 & SB     & 12$^h$05$^m$34.19$^s$ & +50$^d$32$^m$20.5$^s$ & 0.00252 & 13.4 & 10.25 & 9070336 & 4 $\times$ 30 & 2 $\times$ 60 \\
NGC\,4194 & SB+AGN\tablenotemark{c} & 12$^h$14$^m$09.64$^s$ & +54$^d$31$^m$34.6$^s$ & 0.00835 & 40.3 & 11.06 & 3757824 & 4 $\times$ 30 & 2 $\times$ 60 \\
Mrk\,52   & SB     & 12$^h$25$^m$42.67$^s$ & +00$^d$34$^m$20.4$^s$ & 0.0071  & 30.1 & 10.14 & 3753216 & 4 $\times$ 30 & 2 $\times$ 60 \\
NGC\,4676 & SB     & 12$^h$46$^m$10.10$^s$ & +30$^d$43$^m$55.0$^s$ & 0.022   & 94.0 & 10.88 & 9073152 & 8 $\times$ 30 & 6 $\times$ 60 \\
NGC\,4818 & SB     & 12$^h$56$^m$48.90$^s$ & -08$^d$31$^m$31.1$^s$ & 0.00355 &  9.4 & 09.75 & 9071104 & 4 $\times$ 30 & 2 $\times$ 60 \\
NGC\,4945 & SB+AGN\tablenotemark{b} & 13$^h$05$^m$27.48$^s$ & -49$^d$28$^m$05.6$^s$ & 0.00186 &  3.9 & 10.48 & 8769280 & 1 $\times$ 6 & 1 $\times$ 60 \\
Mrk\,266  & SB+AGN\tablenotemark{b} & 13$^h$38$^m$17.69$^s$ & +48$^d$16$^m$33.9$^s$ & 0.02786 &115.8 & 11.49 & 3755264\tablenotemark{d} & 4 $\times$ 30 & 2 $\times$ 60 \\
NGC\,7252 & SB     & 22$^h$20$^m$44.77$^s$ & -24$^d$40$^m$41.8$^s$ & 0.0156  & 66.4 & 10.75 & 9074688 & 4 $\times$ 30 & 3 $\times$ 60 \\
NGC\,7714 & SB     & 23$^h$36$^m$14.10$^s$ & +02$^d$09$^m$18.6$^s$ & 0.00933 & 38.2 & 10.72 & 3756800 & 4 $\times$ 30 & 2 $\times$ 60 \\
\enddata
\tablenotetext{a}{This is the AOR ID of the high resolution observations only.}
\tablenotetext{b}{The evidence for an AGN in these objects come
    from the literature  (NED).}
\tablenotetext{c}{The evidence for an AGN comes from the
  detection of the \ion{[Ne}{5]} line in this work.}
\tablenotetext{d}{The coordinates in the Table for this galaxy
    are slightly off the nucleus (see \S2 and Fig.~1). The galaxy is still
    included in the LH slit but for the smaller SH slit we used an
    observation from a different program (PID$=$3237, AORkey$=$10510592)
    where the source is well centered.}

\end{deluxetable*}

As part of our continuing efforts to provide a comprehensive picture
of the starburst phenomenon at low redshifts, we here present an atlas
of high resolution (R$\sim$600) 9.7-37$\mu$m spectra of well known
nearby starburst galaxies, taken with the Infrared
Spectrograph\footnote{The {\sl IRS} was a collaborative venture
  between Cornell University and Ball Aerospace Corporation funded by
  NASA through the Jet Propulsion Laboratory and the Ames Research
  Center.} \citep{Houck04} on-board Spitzer. This follows on from our
earlier work on the same sample using the low resolution IRS modules
\citep{Brandl06}.  We have created a high S/N template from these
spectra.
  Our goal is to make this set of high quality spectra available
    to the community, and to report on their characteristic features.
  Observations are described in \S\ref{obs}, and data reduction and
  analysis is described in \S\ref{danalysis}. We present results in
  \S\ref{results} and discussion in \S\ref{discussion}.  Finally, we
  summarize our conclusions in \S\ref{conclusion}.

\begin{figure*}
 \begin{center}
 \includegraphics[width=17cm]{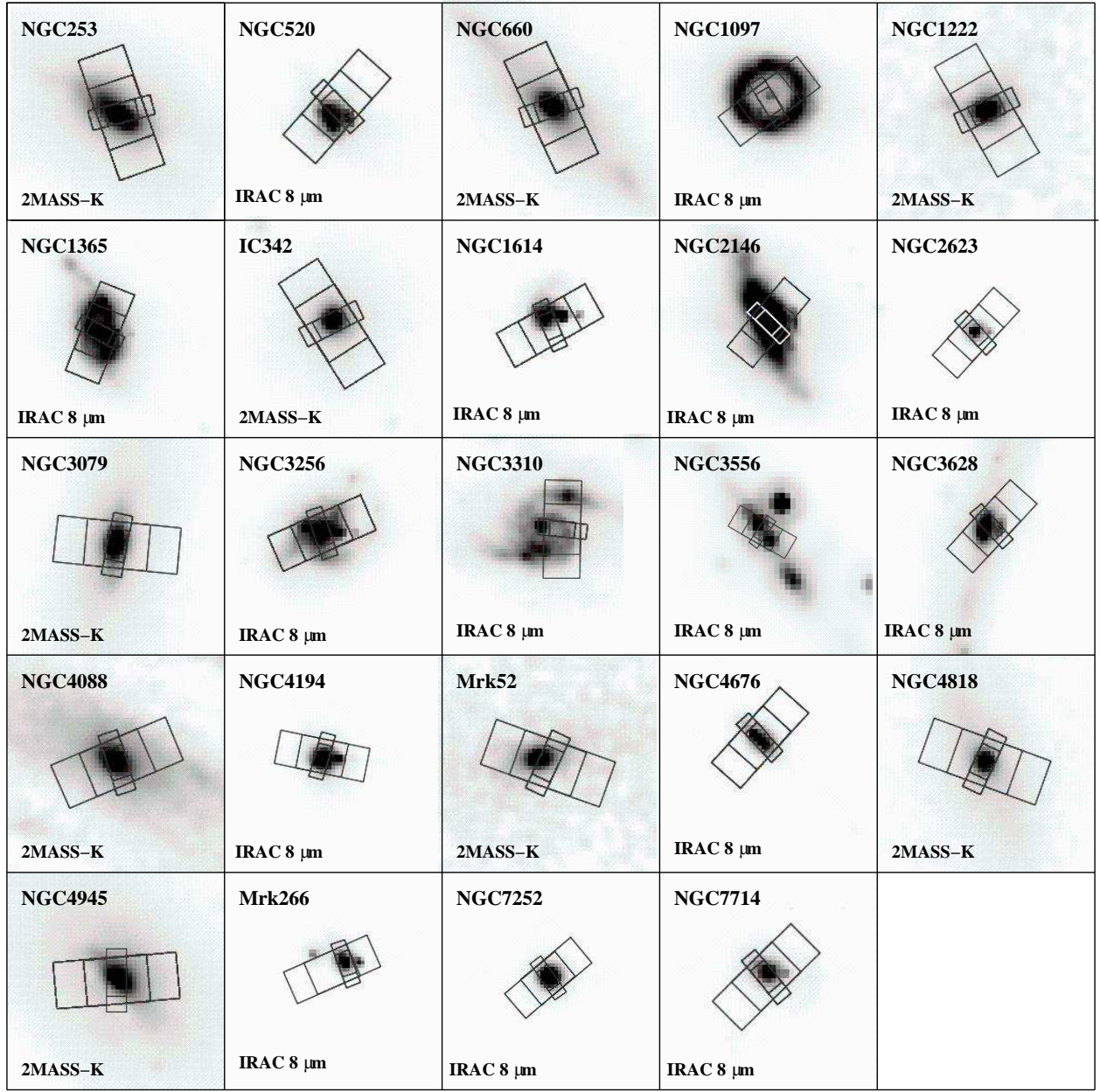}
 \end{center}
 \caption{ IRAC 8~$\mu$m, and 2MASS-K images of the targets with
     the SH and LH slits overlayed on them. The grey scale of the
     image is linear up to the 50\% level of brightness and is black
     above it.}
\end{figure*}

\begin{figure*}\label{allspectra}
 \begin{center}
 \includegraphics[width=17cm]{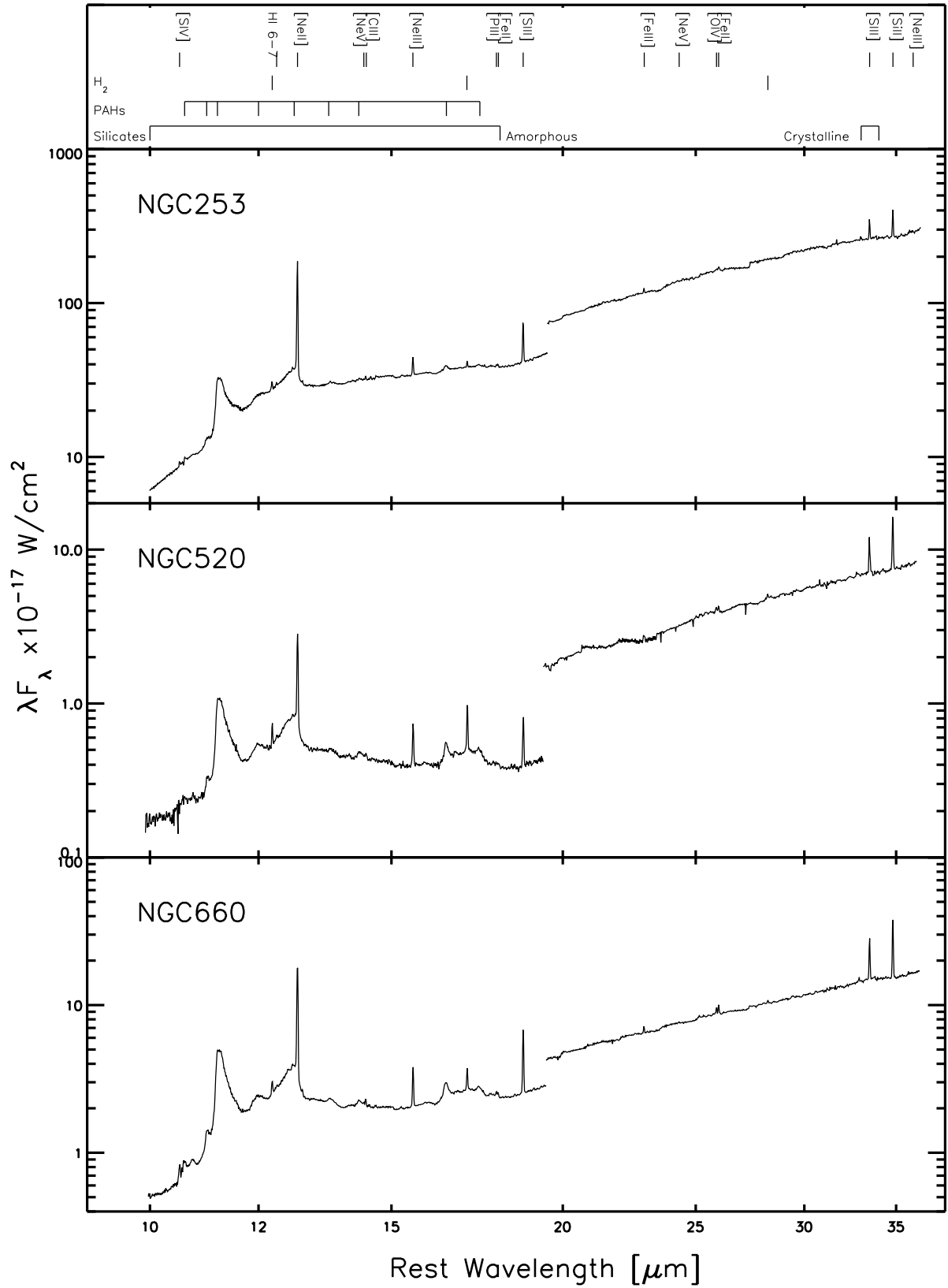}
 \end{center}
 \caption{SH and LH spectra. The main lines and features are indicated
   in the top panel. The discontinuity around 19.5~$\mu$m is caused by
   the extended nature of the sources combined with the different
   sizes of the SH and LH apertures, see also \citet{Brandl06}, their
   Figure~1.}
\end{figure*}

\setcounter{figure}{1}

\begin{figure*}
 \begin{center}
 \includegraphics[width=17cm]{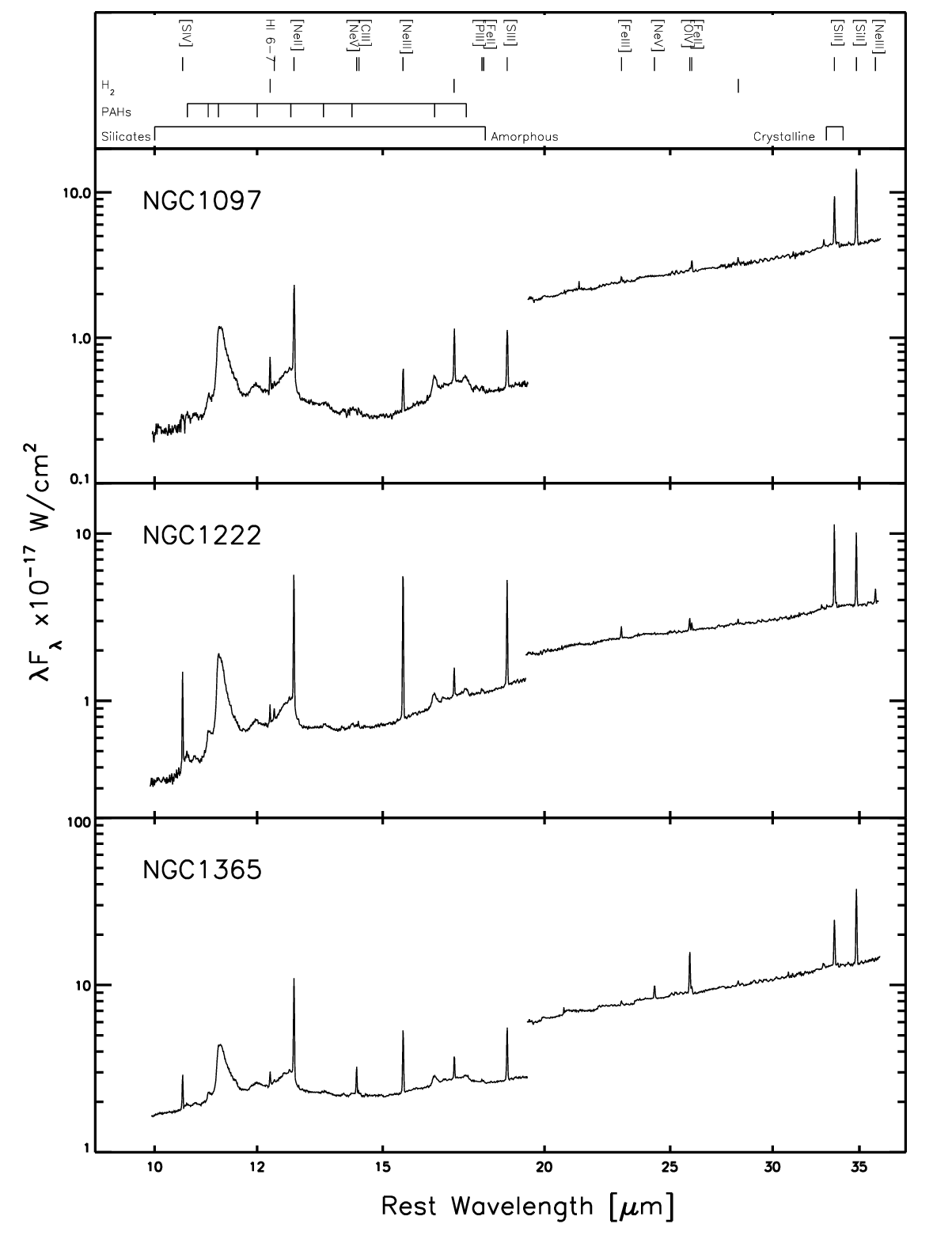}
 \end{center}
 \caption{Continued.}
\end{figure*}

\setcounter{figure}{1}

\begin{figure*}
  \begin{center}
  \includegraphics[width=17cm]{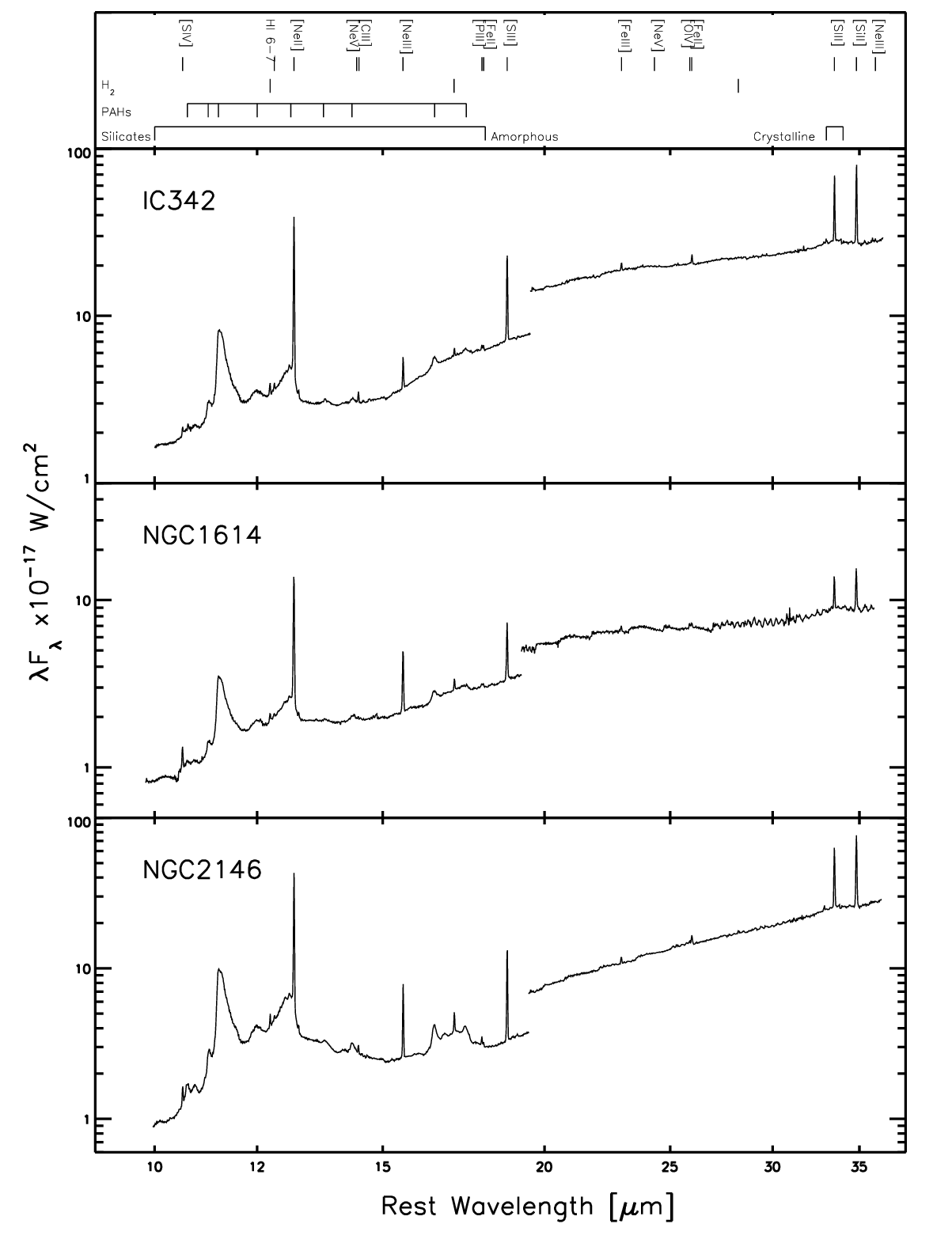}
  \end{center}
  \caption{Continued.}
\end{figure*}

\setcounter{figure}{1}

\begin{figure*}
  \begin{center}
  \includegraphics[width=17cm]{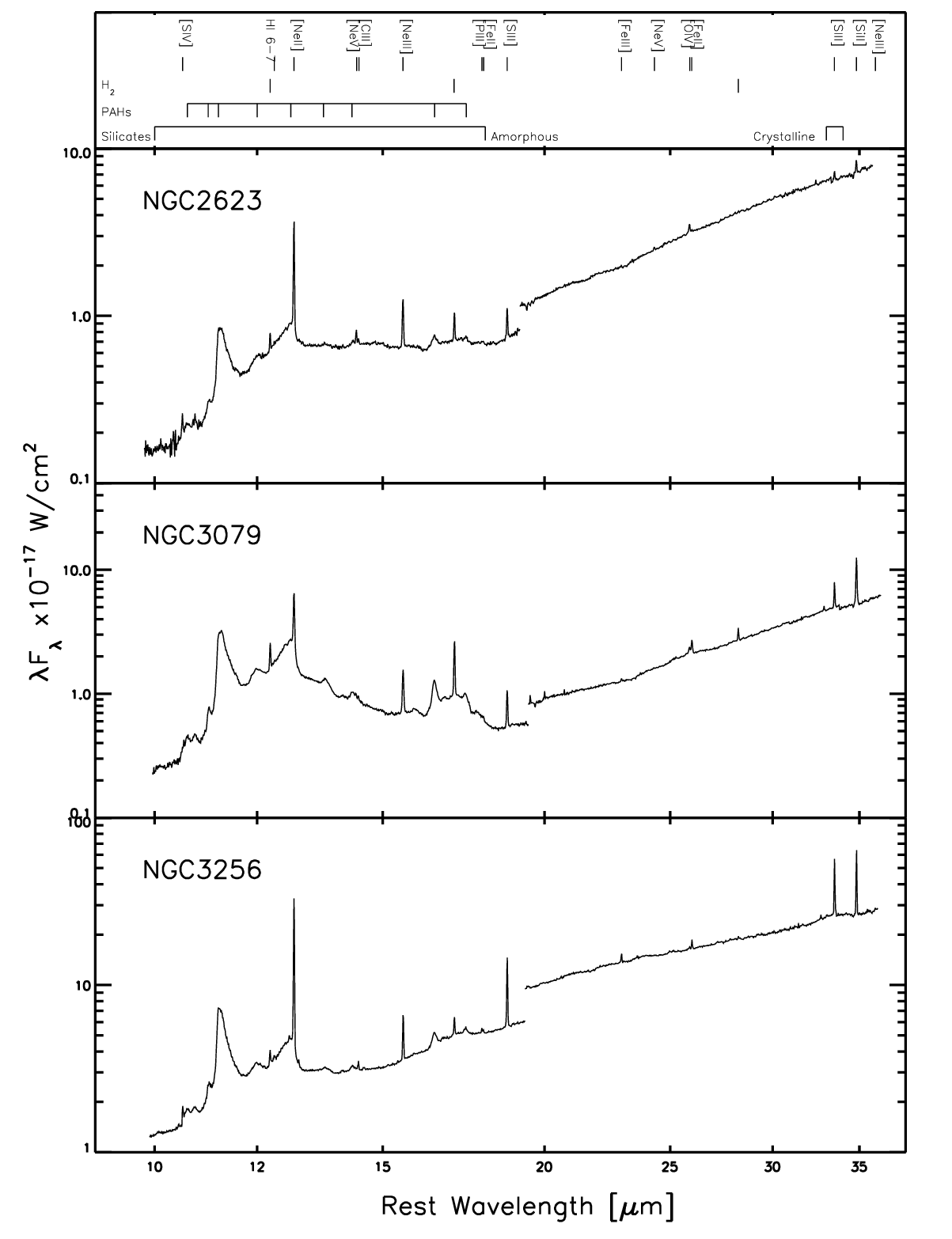}
  \end{center}
  \caption{Continued.}
\end{figure*}

\setcounter{figure}{1}

\begin{figure*}
  \begin{center}
  \includegraphics[width=17cm]{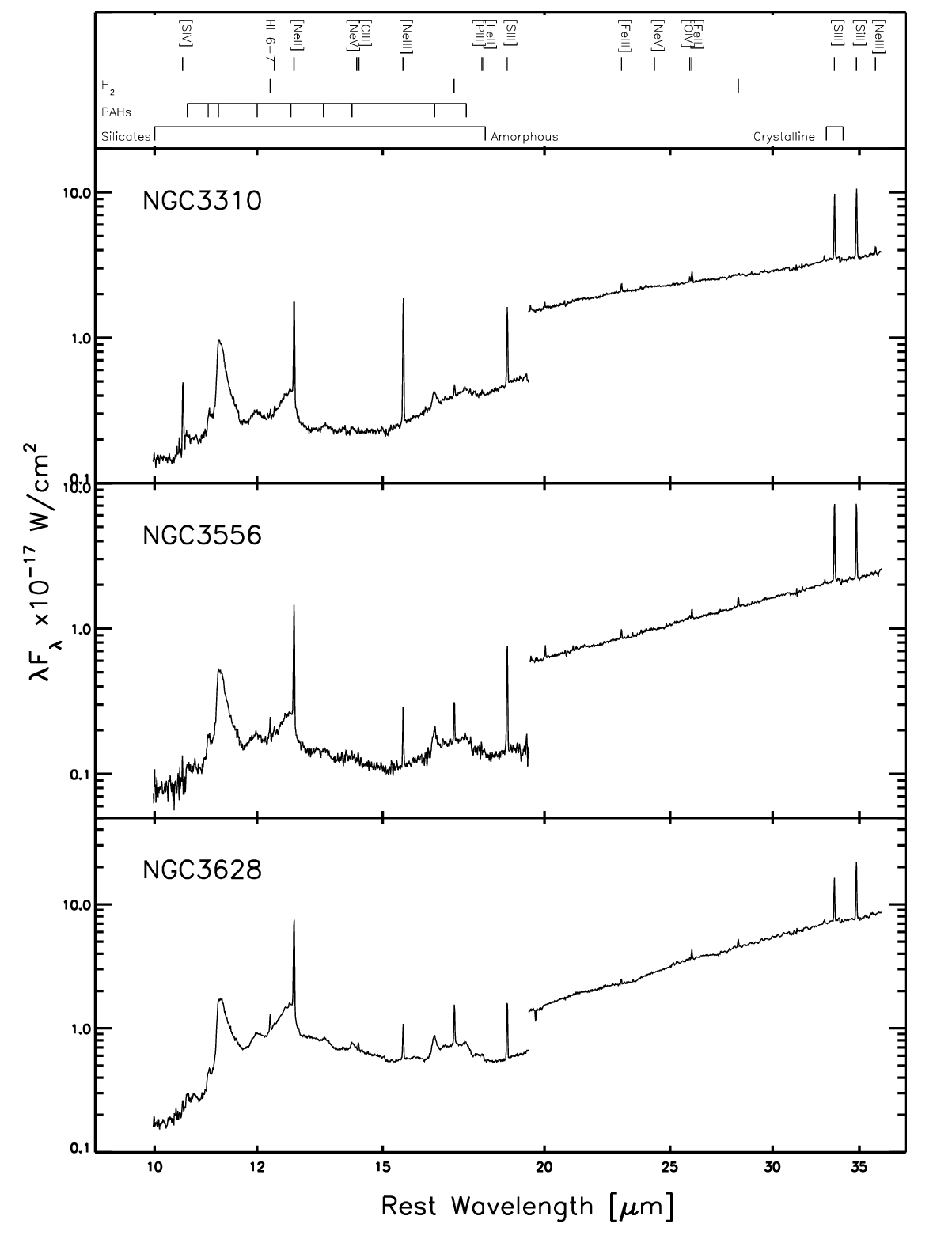}
  \end{center}
  \caption{Continued.}
\end{figure*}

\setcounter{figure}{1}

\begin{figure*}
  \begin{center}
  \includegraphics[width=17cm]{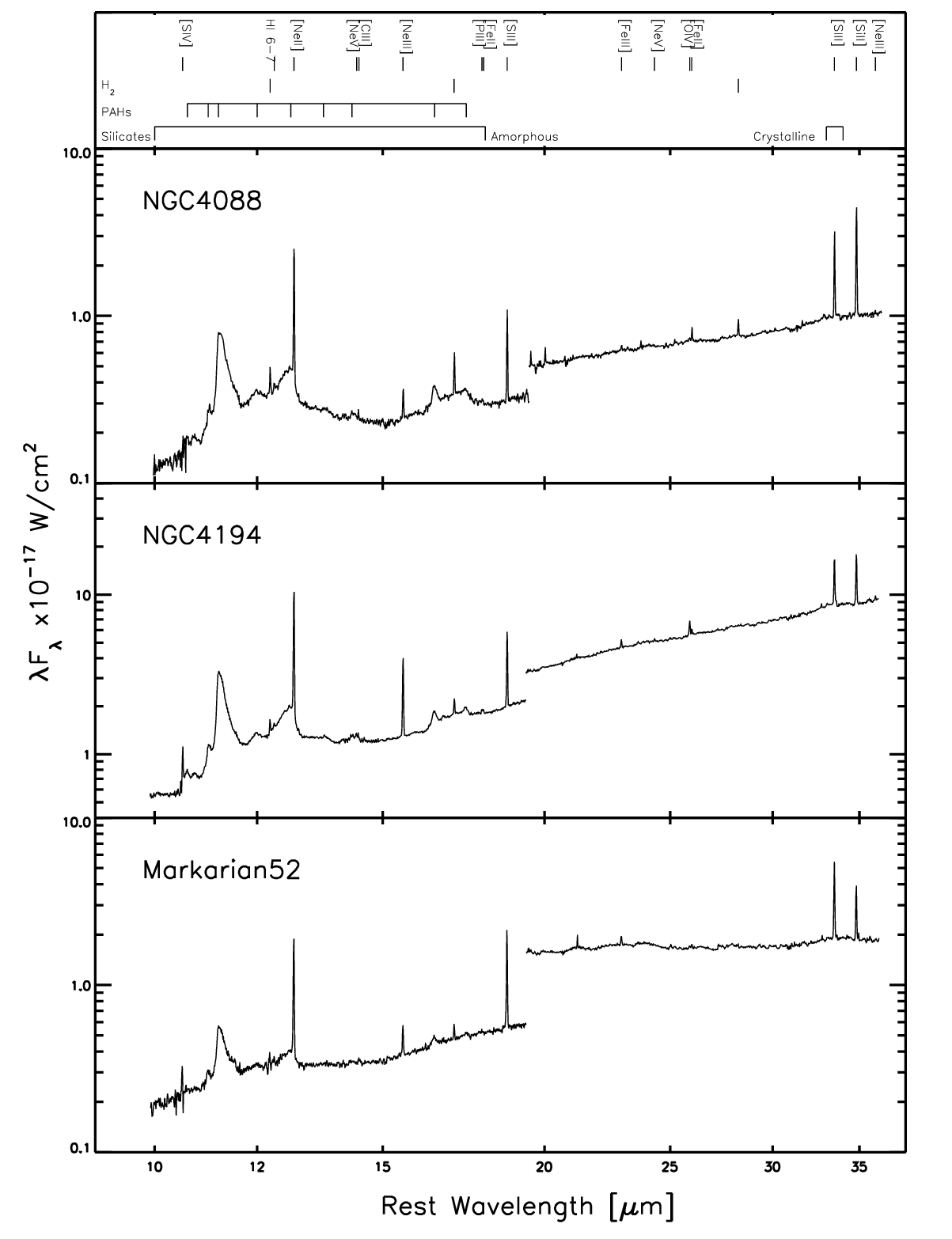}
  \end{center}
  \caption{Continued.}
\end{figure*}

\setcounter{figure}{1}

\begin{figure*}
  \begin{center}
  \includegraphics[width=17cm]{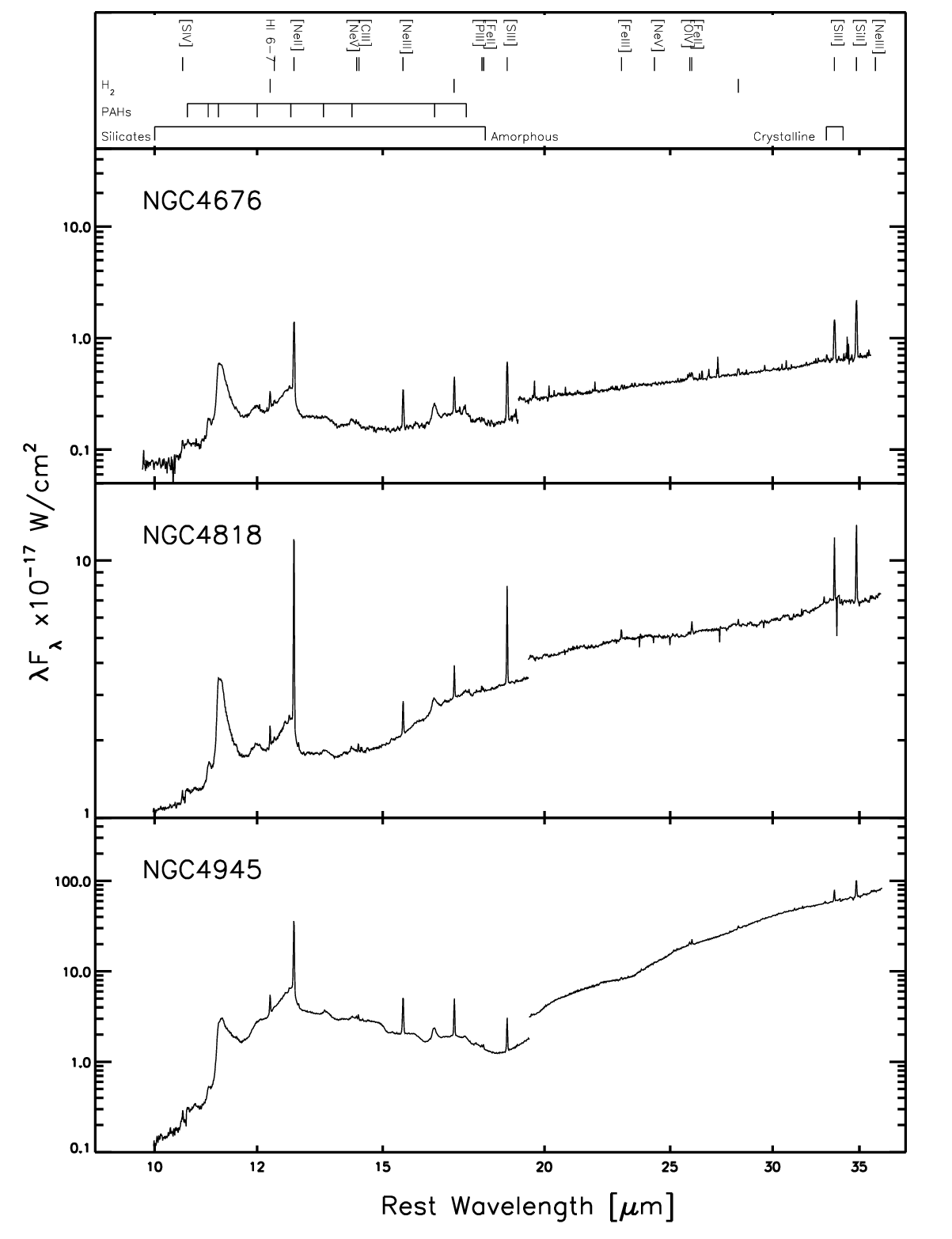}
  \end{center}
  \caption{Continued.}
\end{figure*}

\setcounter{figure}{1}

\begin{figure*}
  \begin{center}
  \includegraphics[width=17cm]{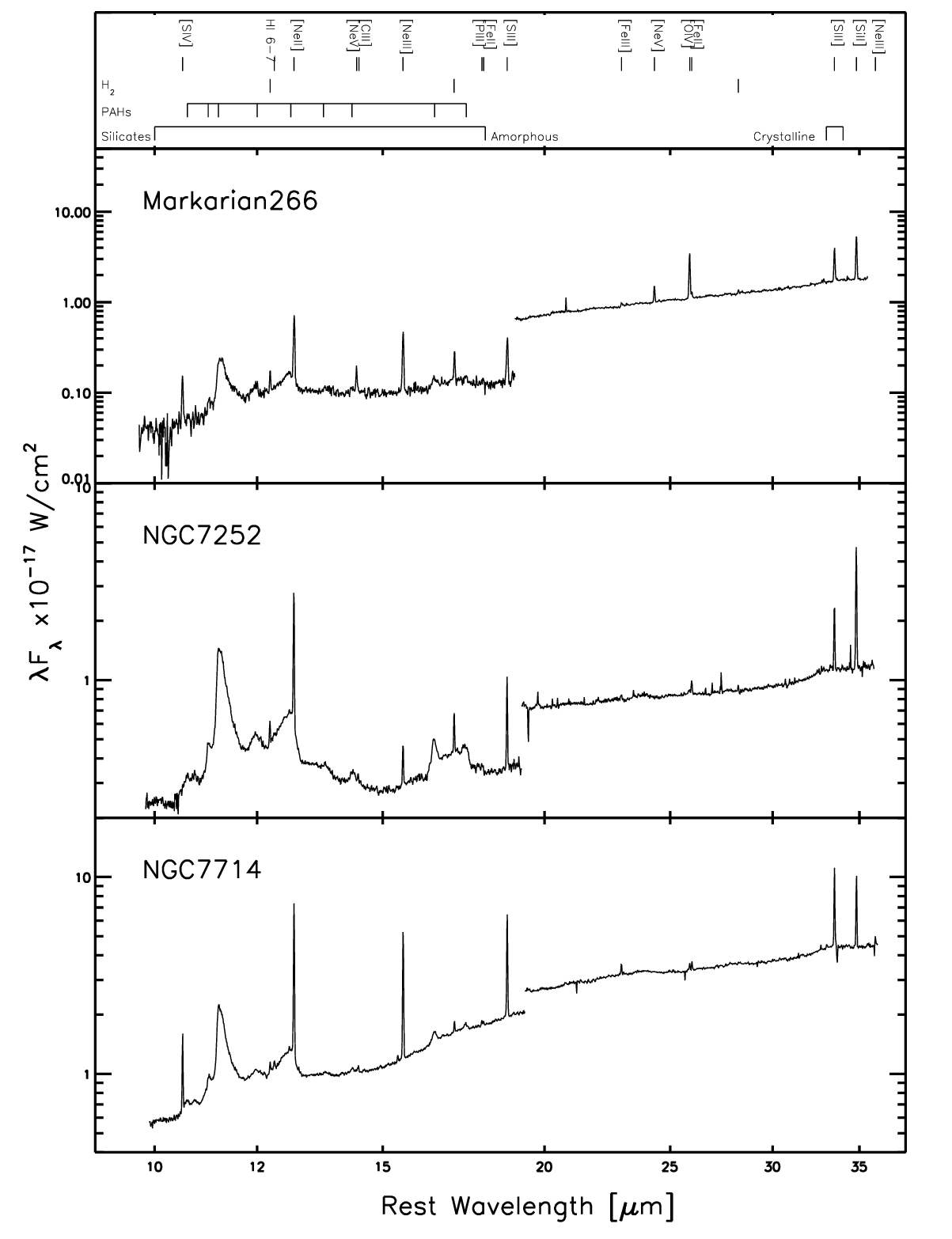}
  \end{center}
  \caption{Continued.}
\end{figure*}

\section{Observations}\label{obs}

Our sample is presented in Table~1 where information on the galaxy
optical classifications, distances, magnitudes and other basic
information on the observations is given.  In nine objects there is
evidence of an active galactic nucleus (AGN), even though it does not
dominate the bolometric luminosity of the system. The sample was
observed as part of the IRS Guaranteed Time program (PID 14) and
consists of 24 well known nearby starburst galaxies, with an average
distance of 33~Mpc and an average infrared (IR) (8-1000$\mu$m)
luminosity of 5$\times10^{10}$ L$_{\odot}$.  The observations were
made using both the high- and low-resolution modules of the IRS. Here
we present the high-resolution spectra. The low-resolution spectra can
be found in the companion paper by \citet{Brandl06}.

The two high-resolution modules, short-high (SH) and long-high( LH),
cover the wavelength region between 10 to 37~$\mu$m at a resolution of
600 and have slit sizes of 4.7\arcsec$\times$11.3\arcsec~and
11.1\arcsec$\times$22.3\arcsec~respectively. The data were taken using
the Staring Mode observing template which produces spectra in two {\em
  nod} positions. The separation between the {\em nod} positions is
3.8\arcsec~for SH and 7.4\arcsec~for LH. The observations were
centered on the nucleus of the galaxies (see Fig.~1) at the
coordinates given in Table~1.  However, as can be seen in Figure~1,
the observations of NGC\,520 and Mrk\,52 are slightly mispointed with
the SH slit measuring part of the nucleus. In NGC\,3310 the
mispointing is more severe and affects both modules and therefore any
information we give about this galaxy in the paper refers to the
position we measured and not the nucleus. Mrk\,266 is also mispointed
but the LH module is large enough to contain the whole galaxy and for
the SH module we made use of an observation of another Spitzer program
in which the source is well centered (PID$=$3237, AORkey$=$10510592).
The area covered by the SH slit is equivalent to a spatial scale of
about 0.085~kpc to 3.2~kpc, depending on the distance to the objects
in our sample. No dedicated background observations were made.  This
means that there will be a small contribution of the zodiacal light to
the spectra in Figures~2, 3 and 4. This has no consequence for the
line measurements. Foreground emission of PAHs from our MW is also
possible but this is much fainter than the targets themselves.  For
information of the mid-infrared continuum properties the reader is
referred to the paper on the low resolution spectra \citep{Brandl06}
where the background contribution to the continuum has been subtracted
from the spectra.

\section{Data Reduction \& Analysis}\label{danalysis}

The data were processed using version s15.3 of the {\em Spitzer}
Science Center's pipeline which is maintained at Cornell and using a
script version of {\em Smart}\footnote{Smart can be downloaded from
  this web site
  http://ssc.spitzer.caltech.edu/archanaly/contributed/smart/index.html.}
\citep{smart}.  The reduction begins from the {\em droop} images
which are equivalent to the more commonly used {\em bcd} data but
without flatfield and stray-cross-light corrections.  Unstable
sensitive pixels (usually referred to as `rogue' pixels) were
replaced using the SSC tool {\em irsclean}\footnote{This tool is
  available from the SSC web site: http://ssc.spitzer.caltech.edu},
which uses campaign-based pixel masks to identify the bad pixels. At
the location of a pixel identified as bad, the empirical profile in
the cross dispersion direction is determined using the average profile
in the neighboring rows. This profile is then scaled to the profile of
the row containing the bad pixel using the remaining good pixels of
that row. The rogue pixels are then replaced by the value given by the
empirical profile. This method of replacing bad pixels is equivalent
to interpolation when the bad pixel is located on the wings of the
spatial profile but it has a definite advantage over interpolation
when the bad pixel is located at the maximum of the spatial profile.

\begin{deluxetable*}{l c c c c c c c c c c c}{!h}
\tabletypesize{\scriptsize}
  \tablewidth{0pt}
  \tablecaption{SH line fluxes\tablenotemark{a}.\label{lines_hr_s}}
  \tablehead{\colhead{Object} & \colhead{\ion{[S}{4]}}  &
    \colhead{H$_2$(S2)}   &  \colhead{HI(7-6)}  &  \colhead{\ion{[Ne}{2]}}
    &     \colhead{\ion{[Ne}{5]}}  &     \colhead{\ion{[Cl}{2]}} &
    \colhead{\ion{[Ne}{3]}}  &    \colhead{H$_2$(S1)}  &
    \colhead{\ion{[P}{3]}}  &     \colhead{\ion{[Fe}{2]}}  &
    \colhead{\ion{[S}{3]}}\\
    & \colhead{10.51\tablenotemark{b}} &   \colhead{12.28} &   \colhead{12.37}  &  \colhead{12.81} &     \colhead{14.32} &   \colhead{14.38}  &   \colhead{15.55}   &  \colhead{17.03}  &   \colhead{17.89}  &   \colhead{17.95}  &  \colhead{18.71}}

  \startdata    

    NGC\,253 &$<$10.50 &   64.32 &$<$27.45 & 2832.33 &$<$20.50 &   25.57 &  204.64 &   67.24 &   22.89 &   25.24 &  666.37 \\
           &    .... &    6.78 &    .... &   64.20 &    .... &    5.92 &    9.59 &    2.35 &    6.01 &    5.97 &   14.93 \\
    NGC\,520 & $<$0.47 &    3.49 & $<$0.65 &   44.62 & $<$0.64 &    0.89 &    7.53 &   10.82 & $<$0.51 & $<$0.48 &    9.10 \\
           &    .... &    0.22 &    .... &    1.25 &    .... &    0.15 &    0.21 &    0.35 &    .... &    .... &    0.59 \\
    NGC\,660 &    2.79 &   12.53 & $<$3.68 &  353.01 & $<$2.26 &    4.05 &   36.96 &   23.52 &    4.89 &    2.52 &   90.02 \\
           &    0.47 &    1.01 &    .... &    7.55 &    .... &    0.72 &    0.49 &    0.90 &    1.00 &    0.36 &    1.71 \\
   NGC\,1097 & $<$0.63 &    6.38 &    0.57 &   37.24 & $<$0.43 & $<$0.47 &    6.33 &   13.22 & $<$0.99 & $<$0.51 &   15.88 \\
           &    .... &    0.50 &    0.11 &    9.27 &    .... &    .... &    0.18 &    0.36 &    .... &    .... &    0.40 \\
   NGC\,1222 &   22.24 &    2.93 &    2.12 &   80.57 & $<$0.57 & $<$0.88 &   89.41 &    9.53 &    1.49 &    0.83 &   64.80 \\
           &    0.45 &    0.16 &    0.20 &    1.30 &    .... &    .... &    1.74 &    0.31 &    0.21 &    0.26 &    1.13 \\
   NGC\,1365 &   23.37 &    9.00 &    3.18 &  139.50 &   18.83 &    2.05 &   59.53 &   18.42 &    2.22 &    1.72 &   53.72 \\
           &    1.42 &    0.91 &    0.39 &    3.27 &    0.51 &    0.37 &    0.92 &    0.63 &    0.38 &    0.38 &    3.58 \\
     IC\,342 &    4.76 &   10.32 &    6.22 &  615.46 & $<$2.41 &    7.99 &   37.20 &   11.69 &    8.27 &    6.55 &  320.03 \\
           &    0.66 &    0.96 &    0.79 &   10.52 &    .... &    0.77 &    0.90 &    0.76 &    0.79 &    0.84 &    5.96 \\
   NGC\,1614 &    6.89 &    5.14 &    1.77 &  249.00 & $<$0.99 &    1.15 &   63.32 &    9.42 &    5.14 &    2.13 &   83.03 \\
           &    0.54 &    0.51 &    0.37 &    7.00 &    .... &    0.32 &    1.59 &    0.58 &    0.58 &    0.51 &    2.64 \\
   NGC\,2146 &    6.30 &   12.79 &    6.81 &  625.00 & $<$2.81 &    5.48 &   91.16 &   26.12 &    7.71 &    3.24 &  190.12 \\
           &    0.44 &    1.35 &    1.53 &   15.35 &    .... &    0.66 &    0.95 &    1.64 &    0.42 &    0.63 &    4.15 \\
   NGC\,2623 &    1.20 &    3.15 & $<$0.49 &   55.55 &    2.71 &    0.69 &   15.08 &    7.27 & $<$0.52 & $<$0.52 &    8.20 \\
           &    0.15 &    0.19 &    .... &    0.66 &    0.15 &    0.13 &    0.24 &    0.23 &    .... &    .... &    0.36 \\
   NGC\,3079 & $<$0.96 &   23.81 & $<$2.09 &   98.96 &    1.31 &    1.57 &   23.17 &   40.63 & $<$0.52 &    0.90 &   11.83 \\
           &    .... &    1.28 &    .... &    2.03 &    0.21 &    0.21 &    0.46 &    1.04 &    .... &    0.13 &    0.16 \\
   NGC\,3256 &    5.25 &   15.04 &    6.19 &  514.19 & $<$2.35 &    5.99 &   64.42 &   27.99 &    6.29 &    3.69 &  171.83 \\
           &    0.57 &    1.07 &    1.38 &    9.34 &    .... &    0.85 &    0.86 &    2.84 &    0.73 &    0.53 &    2.07 \\
   NGC\,3310 &    4.46 & $<$0.56 &    0.58 &   27.57 & $<$0.25 &    0.29 &   28.35 &    1.88 & $<$0.61 & $<$0.59 &   20.50 \\
           &    0.41 &    .... &    0.13 &    0.61 &    .... &    0.10 &    0.45 &    0.11 &    .... &    .... &    0.31 \\
   NGC\,3556 & $<$0.56 &    1.06 &    0.71 &   21.47 & $<$0.37 & $<$0.40 &    3.23 &    2.90 & $<$0.45 & $<$0.34 &   12.36 \\
           &    .... &    0.16 &    0.18 &    0.42 &    .... &    .... &    0.09 &    0.12 &    .... &    .... &    0.22 \\
   NGC\,3628 & $<$0.90 &    6.53 & $<$1.68 &  125.63 &    0.90 &    2.35 &   10.05 &   16.24 & $<$0.79 & $<$0.96 &   21.32 \\
           &    .... &    0.43 &    .... &    2.37 &    0.35 &    0.18 &    0.29 &    0.60 &    .... &    .... &    0.77 \\
   NGC\,4088 & $<$0.67 &    2.70 &    0.95 &   37.03 & $<$0.39 &    0.51 &    2.45 &    4.72 & $<$0.40 & $<$0.37 &   12.50 \\
           &    .... &    0.32 &    0.11 &    1.16 &    .... &    0.16 &    0.14 &    0.16 &    .... &    .... &    0.26 \\
   NGC\,4194 &   11.46 &    5.03 &    2.12 &  165.47 &    2.96 &    2.49 &   53.99 &    9.25 &    1.70 &    1.90 &   70.92 \\
           &    0.75 &    0.51 &    0.58 &    3.90 &    0.65 &    0.38 &    0.78 &    0.66 &    0.32 &    0.38 &    1.44 \\
     Mrk\,52 &    1.02 &    1.50 & $<$0.61 &   29.38 & $<$0.57 & $<$0.43 &    3.82 &    1.80 & $<$0.45 & $<$0.39 &   28.45 \\
           &    0.10 &    0.22 &    .... &    0.53 &    .... &    .... &    0.29 &    0.19 &    .... &    .... &    1.63 \\
   NGC\,4676 &    0.56 &    2.03 &    0.48 &   27.78 & $<$0.25 & $<$0.31 &    4.68 &    5.19 & $<$0.36 & $<$0.36 &   11.32 \\
           &    0.06 &    0.14 &    0.09 &    0.96 &    .... &    .... &    0.16 &    0.32 &    .... &    .... &    0.37 \\
   NGC\,4818 &    2.21 &    5.87 &    1.71 &  184.93 & $<$1.21 &    2.27 &   13.73 &   16.83 &    2.61 &    1.41 &   71.96 \\
           &    0.16 &    0.35 &    0.50 &    5.60 &    .... &    0.35 &    0.45 &    0.57 &    0.35 &    0.18 &    1.05 \\
   NGC\,4945 &    0.72 &   33.93 & $<$3.66 &  583.84 &    3.84 &    6.51 &   68.98 &   54.15 &    1.77 &    2.85 &   30.53 \\
           &    0.18 &    0.85 &    .... &   12.50 &    0.34 &    0.27 &   14.25 &    4.22 &    0.25 &    0.24 &    0.58 \\
    Mrk\,266 &    9.00 &    3.72 &    0.69 &   57.04 &    7.96 &    0.46 &   27.95 &    8.40 &    0.67 &    0.48 &   24.33 \\
           &    0.38 &    0.32 &    0.06 &    1.64 &    0.17 &    0.07 &    0.76 &    0.97 &    0.15 &    0.14 &    0.69 \\
   NGC\,7252 & $<$0.56 &    2.70 & $<$0.57 &   41.68 & $<$0.33 &    0.56 &    3.72 &    5.10 & $<$0.54 & $<$0.57 &   11.96 \\
           &    .... &    0.29 &    .... &    1.94 &    .... &    0.10 &    0.14 &    0.18 &    .... &    .... &    0.31 \\
   NGC\,7714 &   14.76 &    2.85 &    2.18 &  102.55 & $<$1.00 & $<$0.68 &   77.42 &    4.43 &    2.01 &    1.58 &   81.50 \\
           &    0.63 &    0.41 &    0.36 &    1.98 &    .... &    .... &    0.92 &    0.35 &    0.25 &    0.27 &    1.10 \\
Template\tablenotemark{c}    &   14.16 &   13.86 &    5.56 &  452.83 &    .... &    4.58 &  100.00 &   30.83 &    5.74 &    3.30 &  187.97 \\

  \enddata
  \tablenotetext{a}{Per object, the line fluxes are given in the first
    row in units of 10$^{-21}$ W~cm$^{-2}$, with the uncertainties
    shown in the row below. Note that, except for the values of
      the template, the fluxes listed here are not to be directly
      compared (unless scaled) to those in Table~3 because of the
      different aperture used.}
  \tablenotetext{b}{ Rest wavelength in $\mu$m.}

  \tablenotetext{c}{ The line fluxes for the template spectrum are
      relative to F(\ion{[Ne}{3]})$=$100.}
\end{deluxetable*}

For each module the individual observations were averaged together,
and spectra extracted using full aperture extraction. Flux calibration
was performed by dividing the averaged spectra by the spectrum of the
calibration star $\xi$~Dra (extracted in the same way as the target)
and multiplying by its theoretical template \citep[][Sloan et al. in
prep]{coh03}.  Fringes, when present, were removed using the IRS
de-fringing tool {\em irsfringe}\footnote{This contributed software
  can be downloaded from
  http://ssc.spitzer.caltech.edu/archanaly/contributed/irsfringe/}.
Finally, the spectra for each {\em nod} position were combined. It is
worth mentioning that in six of the starbursts (IC~342, NGC\,520,
NGC\,1097, NGC\,1614, NGC\,3310, and NGC\,3628) the continua differed between
{\em nod} positions. These sources are extended (see Figure~1),
and this difference may reflect the different regions the {\em nod}
positions are sampling. Since we are interested in
the overall properties of the emission lines and PAH bands, and
differences in these quantities lie within the errors in each nod
position, we combined the nod positions in these starbursts as it was
done for the rest of the sample.

\begin{deluxetable*}{l c c c c c c c c}
\tabletypesize{\scriptsize}
  \tablewidth{0pt}
  \tablecaption{LH line fluxes\tablenotemark{a}.\label{lines_hr_t}}
  \tablehead{\colhead{Object} & \colhead{\ion{[Fe}{3]}}  &    \colhead{\ion{[Ne}{5]}}  &    \colhead{\ion{[O}{4]}}  &    \colhead{\ion{[Fe}{2]}}  &    \colhead{H$_2$(S0)} &   \colhead{\ion{[S}{3]}} &     \colhead{\ion{[Si}{2]}} &      \colhead{\ion{[Ne}{3]}}\\
    &  \colhead{22.93\tablenotemark{b}}  &    \colhead{24.31} &   \colhead{25.89} &   \colhead{25.98}  &    \colhead{28.22} &   \colhead{33.48}  &   \colhead{34.81}   &   \colhead{36.01}}
  \startdata

    NGC\,253 &  120.84 &$<$73.36 &  154.74 &  250.56 &$<$77.55 & 1538.03 & 2412.03 & -243.86 \\
           &   18.17 &    .... &   26.95 &   24.43 &    .... &   30.06 &   48.02 &    .... \\
    NGC\,520 &    5.95 & $<$1.42 &    8.10 &    7.56 &    8.23 &   89.44 &  190.71 & $<$6.90 \\
           &    0.67 &    .... &    0.88 &    0.78 &    0.60 &    8.60 &    3.41 &    .... \\
    NGC\,660 &   12.33 & $<$6.59 &   18.80 &   22.92 &   10.58 &  246.07 &  441.96 & $<$8.96 \\
           &    1.64 &    .... &    2.57 &    1.82 &    2.67 &    3.21 &    3.56 &    .... \\
   NGC\,1097 &    4.20 & $<$1.07 &    4.41 &   11.48 &    6.59 &  100.05 &  225.19 & $<$5.29 \\
           &    0.18 &    .... &    1.21 &    1.07 &    0.69 &    3.95 &    7.00 &    .... \\
   NGC\,1222 &    7.04 & $<$0.78 &    9.92 &    4.74 &    3.25 &  132.52 &  112.05 &   12.72 \\
           &    0.38 &    .... &    0.48 &    0.38 &    0.34 &    2.30 &    1.70 &    1.30 \\
   NGC\,1365 &    9.24 &   35.96 &  141.57 &   22.20 &   15.91 &  246.89 &  500.30 &$<$14.78 \\
           &    1.20 &    1.18 &    2.06 &    2.33 &    1.51 &   12.17 &   17.78 &    .... \\
     IC\,342 &   36.43 & $<$4.95 & $<$7.70 &   51.87 & $<$9.10 &  672.46 &  985.73 &$<$29.96 \\
           &    2.69 &    .... &    .... &    1.89 &    .... &    7.28 &   10.29 &    .... \\
   NGC\,1614 &   13.14 & $<$3.27 &    8.68 &   12.99 & $<$8.98 &  101.06 &  148.60 &   12.15 \\
           &    2.10 &    .... &    0.85 &    1.15 &    .... &    2.11 &    4.11 &    3.44 \\
   NGC\,2146 &   23.05 & $<$3.70 &   19.33 &   52.69 &   16.66 &  848.02 & 1209.35 &$<$60.59 \\
           &    2.58 &    .... &    4.48 &    4.59 &    2.33 &   36.93 &   21.70 &    .... \\
   NGC\,2623 & $<$1.31 &    2.21 &    9.62 &    2.23 & $<$2.08 &   13.78 &   28.74 & $<$9.41 \\
           &    .... &    0.14 &    0.72 &    0.51 &    .... &    1.41 &    1.89 &    .... \\
   NGC\,3079 &    1.54 & $<$0.91 &    8.45 &   14.57 &   11.02 &   56.88 &  173.51 & $<$7.05 \\
           &    0.21 &    .... &    0.84 &    0.62 &    0.75 &    2.70 &    5.83 &    .... \\
   NGC\,3256 &   31.55 & $<$3.85 &   12.23 &   33.56 &   12.70 &  484.64 &  623.37 &$<$33.15 \\
           &    2.77 &    .... &    3.02 &    1.82 &    1.32 &   13.79 &    9.81 &    .... \\
   NGC\,3310 &    5.66 & $<$1.00 &    4.03 &    8.78 & $<$2.32 &  106.02 &  143.54 &    8.01 \\
           &    0.59 &    .... &    0.64 &    0.74 &    .... &    1.17 &    1.58 &    1.19 \\
   NGC\,3556 &    2.24 & $<$1.18 & $<$1.50 &    3.14 &    4.55 &   96.37 &   96.44 & $<$3.00 \\
           &    0.34 &    .... &    .... &    0.47 &    0.47 &    1.18 &    0.80 &    .... \\
   NGC\,3628 &    3.69 & $<$1.03 & $<$2.37 &   11.96 &   11.57 &  149.93 &  277.55 &$<$10.09 \\
           &    0.60 &    .... &    .... &    1.07 &    0.42 &    2.20 &    7.75 &    .... \\
   NGC\,4088 &    0.98 & $<$0.50 &    0.73 &    2.15 &    3.38 &   35.02 &   63.33 & $<$1.33 \\
           &    0.09 &    .... &    0.21 &    0.27 &    0.23 &    0.41 &    0.73 &    .... \\
   NGC\,4194 &    9.96 &    4.04 &   27.45 &    8.48 & $<$2.66 &  151.01 &  185.82 &$<$10.83 \\
           &    1.00 &    0.23 &    1.63 &    0.96 &    .... &    1.65 &    3.31 &    .... \\
     Mrk\,52 &    4.53 & $<$0.83 &    1.31 &    1.90 & $<$1.35 &   60.43 &   38.02 & $<$2.44 \\
           &    0.61 &    .... &    0.23 &    0.31 &    .... &    1.07 &    0.60 &    .... \\
   NGC\,4676 & $<$0.58 & $<$0.33 &    1.37 &    1.78 &    1.89 &   23.95 &   41.27 & $<$2.49 \\
           &    .... &    .... &    0.23 &    0.22 &    0.25 &    1.51 &    1.42 &    .... \\
   NGC\,4818 &    7.96 & $<$1.96 & $<$1.59 &   10.83 &    4.68 &   80.54 &  128.66 & $<$5.61 \\
           &    0.50 &    .... &    .... &    0.88 &    0.51 &    2.44 &    1.97 &    .... \\
   NGC\,4945 &    8.69 & $<$7.95 &   47.85 &   55.21 &   44.56 &  338.91 &  846.23 & -160.25 \\
           &    2.09 &    .... &    6.95 &    5.35 &    8.72 &   70.93 &   50.98 &    .... \\
    Mrk\,266 &    2.15 &   11.12 &   52.94 &    4.35 & $<$1.60 &   50.93 &   87.02 &    .... \\
           &    0.29 &    0.59 &    5.66 &    0.70 &    .... &    2.91 &    4.44 &    .... \\
   NGC\,7252 &    1.36 & $<$0.51 &    1.33 &    3.12 &    1.10 &   24.75 &   70.53 & $<$2.20 \\
           &    0.26 &    .... &    0.25 &    0.20 &    0.33 &    0.47 &    0.43 &    .... \\
   NGC\,7714 &    7.88 & $<$1.77 &    5.51 &    5.93 & $<$1.84 &  115.90 &  102.97 &    7.45 \\
           &    0.70 &    .... &    0.94 &    0.54 &    .... &    2.45 &    1.63 &    1.29 \\
 Template\tablenotemark{c}    &   11.25 &    .... &    7.85 &   16.07 &    9.51 &  267.16 &  364.23 &   8.10 \\

  \enddata
  \tablenotetext{a}{For each object, the line fluxes are given in the
    first row in units of 10$^{-21}$ W~cm$^{-2}$, with the
    uncertainties given in the row below. Note that the fluxes
      listed here are not to be directly compared (unless scaled) to
      those in Table~2 because of the different aperture used.}
  \tablenotetext{b}{Rest wavelength in $\mu$m.}
  \tablenotetext{c}{ The line fluxes for the template spectrum are
      relative to F(\ion{[Ne}{3]})$=$100.}

\end{deluxetable*}

\begin{deluxetable*}{lccccccccccc}
\tabletypesize{\scriptsize}
  \tablewidth{0pt}
  \tablecaption{PAH fluxes\tablenotemark{a b}.\label{lines_hr_u}}
  \tablehead{\colhead{Object} &    \colhead{F(10.6)\tablenotemark{c}}  &   \colhead{F(10.7)\tablenotemark{d}}  &    \colhead{F(11.0)} &    \colhead{F(11.3)}   &    \colhead{F(12.0)}  &   \colhead{F(12.74)} &     \colhead{F(13.5)}   &    \colhead{F(14.2)}   &   \colhead{F(16.45)}  &   \colhead{F(16.74)} &    \colhead{F(17.0)}}
  \startdata

      NGC\,253 &    .... &    3.39 &   19.00 &  405.27 &   18.67 &  257.50 &   19.85 &   13.78 &   23.67 &    .... &   15.86 \\
             &    .... &    0.35 &    1.24 &   20.26 &    2.30 &   12.88 &    2.16 &    3.30 &    4.98 &    .... &    6.98 \\
      NGC\,520 &    0.23 &    0.08 &    0.91 &   17.21 &    1.11 &    9.66 &    0.34 &    0.32 &    0.86 &    0.03 &    1.36 \\
             &    0.02 &    0.02 &    0.05 &    0.86 &    0.06 &    0.48 &    0.06 &    0.02 &    0.13 &    0.01 &    0.07 \\
      NGC\,660 &    0.52 &    0.77 &    6.07 &   83.40 &    4.60 &   43.45 &    1.05 &    2.09 &    5.64 &    .... &    4.06 \\
             &    0.21 &    0.10 &    0.30 &    4.17 &    0.44 &    2.17 &    0.26 &    0.22 &    0.32 &    .... &    0.38 \\
     NGC\,1097 &    0.30 &    .... &    1.17 &   18.84 &    0.99 &    6.32 &    0.29 &    0.32 &    1.06 &    .... &    0.99 \\
             &    0.02 &    .... &    0.06 &    0.94 &    0.05 &    0.32 &    0.07 &    0.02 &    0.23 &    .... &    0.05 \\
     NGC\,1222 &    0.66 &    0.26 &    2.72 &   27.56 &    1.09 &    9.69 &    0.38 &    0.27 &    1.54 &    0.33 &    1.53 \\
             &    0.12 &    0.08 &    0.14 &    1.38 &    0.18 &    0.48 &    0.10 &    0.09 &    0.17 &    0.13 &    0.22 \\
     NGC\,1365 &    0.70 &    0.51 &    3.64 &   51.38 &    2.45 &   20.04 &    1.14 &    0.65 &    2.64 &    0.31 &    3.20 \\
             &    0.50 &    0.28 &    0.46 &    2.57 &    0.49 &    1.08 &    0.23 &    0.27 &    0.32 &    0.23 &    0.46 \\
       IC\,342 &    1.41 &    0.64 &    9.70 &  119.19 &    6.38 &   47.77 &    3.04 &    3.26 &    6.21 &    .... &    4.67 \\
             &    0.16 &    0.11 &    0.49 &    5.96 &    0.32 &    2.39 &    0.15 &    0.16 &    0.45 &    .... &    0.23 \\
     NGC\,1614 &    1.18 &    0.12 &    3.82 &   47.29 &    2.46 &   22.80 &    0.30 &    1.98 &    3.35 &    0.34 &    3.93 \\
             &    0.06 &    0.03 &    0.19 &    2.36 &    0.12 &    1.14 &    0.11 &    0.10 &    0.60 &    0.03 &    0.20 \\
     NGC\,2146 &    1.38 &    1.36 &   14.80 &  168.24 &    7.03 &   84.87 &    1.58 &    4.69 &    9.94 &    1.19 &   12.16 \\
             &    0.21 &    0.13 &    0.74 &    8.41 &    0.41 &    4.24 &    0.31 &    0.36 &    0.50 &    0.36 &    0.75 \\
     NGC\,2623 &    0.29 &    0.15 &    0.91 &   12.61 &    0.63 &    7.37 &    0.43 &    0.40 &    1.13 &    .... &    0.74 \\
             &    0.04 &    0.02 &    0.05 &    0.63 &    0.06 &    0.37 &    0.03 &    0.03 &    0.06 &    .... &    0.07 \\
     NGC\,3079 &    .... &    .... &    3.22 &   56.68 &    3.13 &   35.39 &    1.58 &    1.59 &    5.07 &    0.24 &    3.41 \\
             &    .... &    .... &    0.16 &    2.83 &    0.16 &    1.77 &    0.08 &    0.08 &    0.25 &    0.02 &    0.17 \\
     NGC\,3256 &    1.31 &    0.86 &    8.65 &  108.19 &    5.41 &   50.18 &    1.50 &    1.69 &    7.77 &    0.99 &    6.27 \\
             &    0.31 &    0.17 &    0.43 &    5.41 &    0.30 &    2.51 &    0.12 &    0.13 &    0.43 &    0.29 &    0.55 \\
     NGC\,3310 &    .... &    .... &    1.21 &   14.79 &    0.48 &    5.03 &    0.44 &    0.31 &    0.78 &    0.08 &    0.74 \\
             &    .... &    .... &    0.06 &    0.74 &    0.02 &    0.25 &    0.02 &    0.02 &    0.10 &    0.01 &    0.04 \\
     NGC\,3556 &    .... &    .... &    0.84 &    8.46 &    0.46 &    3.17 &    0.31 &    0.29 &    0.44 &    .... &    0.44 \\
             &    .... &    .... &    0.04 &    0.42 &    0.02 &    0.16 &    0.05 &    0.01 &    0.11 &    .... &    0.03 \\
     NGC\,3628 &    0.02 &    0.23 &    1.81 &   29.98 &    1.48 &   19.88 &    0.01 &    0.63 &    2.04 &    0.19 &    2.38 \\
             &    0.02 &    0.01 &    0.09 &    1.50 &    0.07 &    0.99 &    0.14 &    0.03 &    0.27 &    0.01 &    0.12 \\
     NGC\,4088 &    .... &    .... &    1.21 &   12.64 &    0.66 &    5.09 &    0.17 &    0.22 &    0.93 &    .... &    0.68 \\
             &    .... &    .... &    0.06 &    0.63 &    0.03 &    0.25 &    0.01 &    0.02 &    0.09 &    .... &    0.03 \\
     NGC\,4194 &    0.88 &    0.27 &    4.84 &   49.18 &    1.55 &   20.15 &    .... &    1.00 &    2.83 &    0.31 &    3.19 \\
             &    0.14 &    0.09 &    0.24 &    2.46 &    0.21 &    1.01 &    .... &    0.16 &    0.19 &    0.15 &    0.27 \\
       Mrk\,52 &    .... &    .... &    0.64 &    5.88 &    0.43 &    2.49 &    .... &    0.17 &    0.35 &    .... &    0.31 \\
             &    .... &    .... &    0.07 &    0.29 &    0.08 &    0.12 &    .... &    0.03 &    0.04 &    .... &    0.05 \\
     NGC\,4676 &    0.23 &    .... &    0.79 &    9.41 &    0.35 &    4.18 &    0.09 &    0.43 &    0.68 &    0.05 &    0.68 \\
             &    0.02 &    .... &    0.04 &    0.47 &    0.03 &    0.21 &    0.02 &    0.02 &    0.03 &    0.02 &    0.06 \\
     NGC\,4818 &    0.59 &    0.48 &    3.72 &   43.53 &    2.07 &   18.95 &    0.55 &    1.18 &    3.14 &    0.41 &    1.87 \\
             &    0.11 &    0.03 &    0.19 &    2.18 &    0.12 &    0.95 &    0.09 &    0.06 &    0.16 &    0.10 &    0.17 \\
     NGC\,4945 &    .... &    .... &    1.36 &   54.64 &    3.64 &   83.84 &    4.85 &    2.64 &    6.20 &    0.17 &    3.49 \\
             &    .... &    .... &    0.13 &    2.73 &    0.57 &    4.19 &    0.24 &    0.13 &    0.31 &    0.06 &    0.17 \\
      Mrk\,266 &    .... &    .... &    0.27 &    3.71 &    0.27 &    1.07 &    .... &    0.19 &    0.23 &    .... &    0.34 \\
             &    .... &    .... &    0.06 &    0.19 &    0.11 &    0.22 &    .... &    0.06 &    0.06 &    .... &    0.13 \\
     NGC\,7252 &    0.14 &    .... &    2.05 &   23.54 &    1.12 &    7.55 &    .... &    0.50 &    1.31 &    .... &    1.54 \\
             &    0.04 &    .... &    0.10 &    1.18 &    0.06 &    0.38 &    .... &    0.05 &    0.07 &    .... &    0.08 \\
     NGC\,7714 &    0.52 &    0.27 &    3.18 &   28.21 &    1.53 &   11.34 &    0.22 &    0.50 &    1.83 &    0.21 &    1.81 \\
             &    0.14 &    0.05 &    0.16 &    1.41 &    0.15 &    0.57 &    0.07 &    0.09 &    0.11 &    0.09 &    0.16 \\
    Template\tablenotemark{e} &    1.24 &    0.59 &    8.57 &  100.00 &    4.65 &   41.26 &    0.61 &    1.31 &    6.58 &    0.47 &    5.04 \\

   \enddata
   \tablenotetext{a}{PAH fluxes and uncertainties (in the row below
     the measurement) in units of 10$^{-20}$W~cm$^{-2}$.}
   \tablenotetext{b}{The anchor points  for the continuum
     were selected at these positions: 10.21, 10.38, 10.64, 10.83,
     11.74, 12.21, 13.08, 13.47$\mu$m for PAHs between 10 and 13$\mu$m;
     13.01, 13.29, 14.55, 14.70$\mu$m for PAHs between 13-15$\mu$m; and
     15.64, 15.83, 16.10, 16.61, 16.90, 17.70$\mu$m for PAHs between
     16-18$\mu$m. The integration limits for the PAHs (in
     the same order as they are in the table) are: 10.42-10.66,
     10.66-10.83, 10.85-11.1, 11.1-11.7, 11.74-12.20, 12.20-13.1,
     13.4-13.68, 13.9-14.1, 16.2-16.62, 16.62-16.88, and 16.9,17.55$\mu$m.}
 
   \tablenotetext{c}{PAH position in parenthesis in $\mu$m.}

   \tablenotetext{b}{The identification of this feature is not know,
     but for simplicity we fold its measurement into this table.}

   \tablenotetext{e}{The  fluxes of PAH features in the template
     have been normalized so that
     F(11.3)$=$100.}

 \end{deluxetable*}

Fine-structure line fluxes are derived by fitting a Gaussian profile
and integrating the flux above a local continuum. A polynomial of
  order one was used to fit the local continuum for a given line,
  except for the \ion{[Ne}{2]} line which sits at the sharp flank of a
  PAH feature (see Fig.~3). For this line, a third order polynomial
  was used to fit the PAH underneath the line. This is done for each
{\em nod} separately and then the flux in both measurements is
averaged.  Errors bars were estimated using the difference between the
two {\em nod} positions or the standard deviation of the residuals of
the fit, whichever is the larger. Upper-limits are derived by
measuring the flux of a Gaussian with a height three times the local
{\em rms}, and with a {\em FWHM} equal to the instrumental resolution.
Note that, except for the line measurements of the template, Tables~2
and 3 quote the line fluxes without any normalization applied to
correct for the difference in aperture between the SH and LH modules.

Following \citet{hon01,ver02};and \citet{pee02}, we have measured the
strength of the PAH emission features by integrating the flux of the
PAH bands above an adopted spline interpolated local continuum. For a
given band the anchor points used to define the continuum (see
Table~4) are the same for all sources, as are the integration limits.
Spectral lines superimposed on the PAH bands were removed before
integration.  Our spline method for measuring PAH fluxes differs from
that of {\em PAHFIT} \citep{Smith07} and others
\citep[e.g.][]{bou98,li01} by adopting a continuum (which ignores the
plateau) and integrating under the bands, rather than assuming a PAH
band profile and performing a simultaneous fit to all the bands.
\citet{gal08} concluded that both methods reveal similar trends in the
data. The uncertainties in the PAH fluxes (Table~4) are based on the
differences in the fluxes measured from the two individual {\em nod}
spectra. However, this does not take into account all the
systematic uncertainties (especially flux calibration). We estimate
these to be 5\% and choose it as our minimum uncertainty.


\begin{figure*}\label{composite}
  \begin{center}
  \includegraphics[width=16cm]{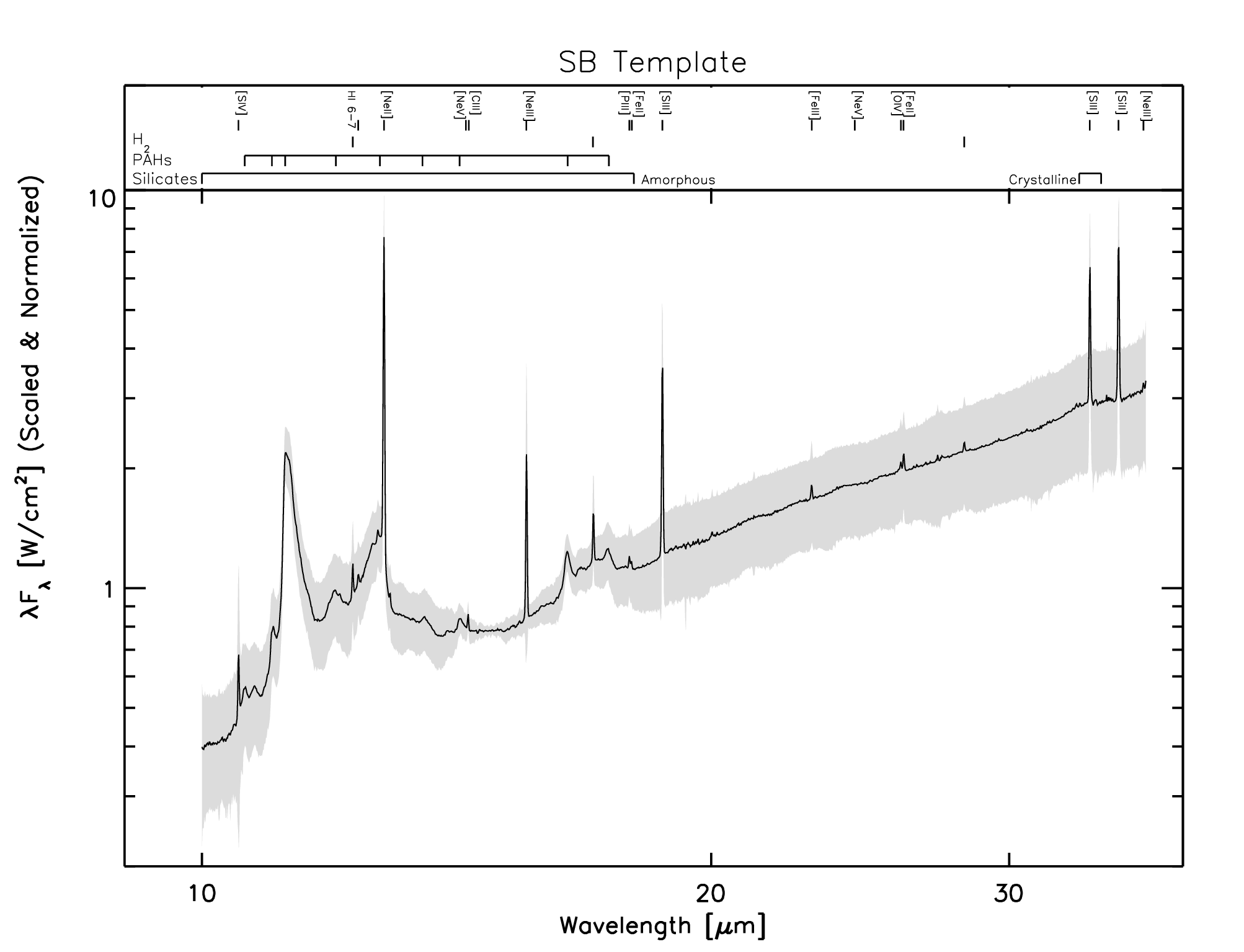}
  \end{center}
  \caption{Template spectrum created by combining the Starburst
    galaxies in our sample.  The grey area
      is the standard deviation in the mean indicating the spectral
      diversity of the starburst sample.}
\end{figure*}

\begin{figure*}\label{compositezoom}
  \begin{center}
  \includegraphics[width=12cm,angle=90]{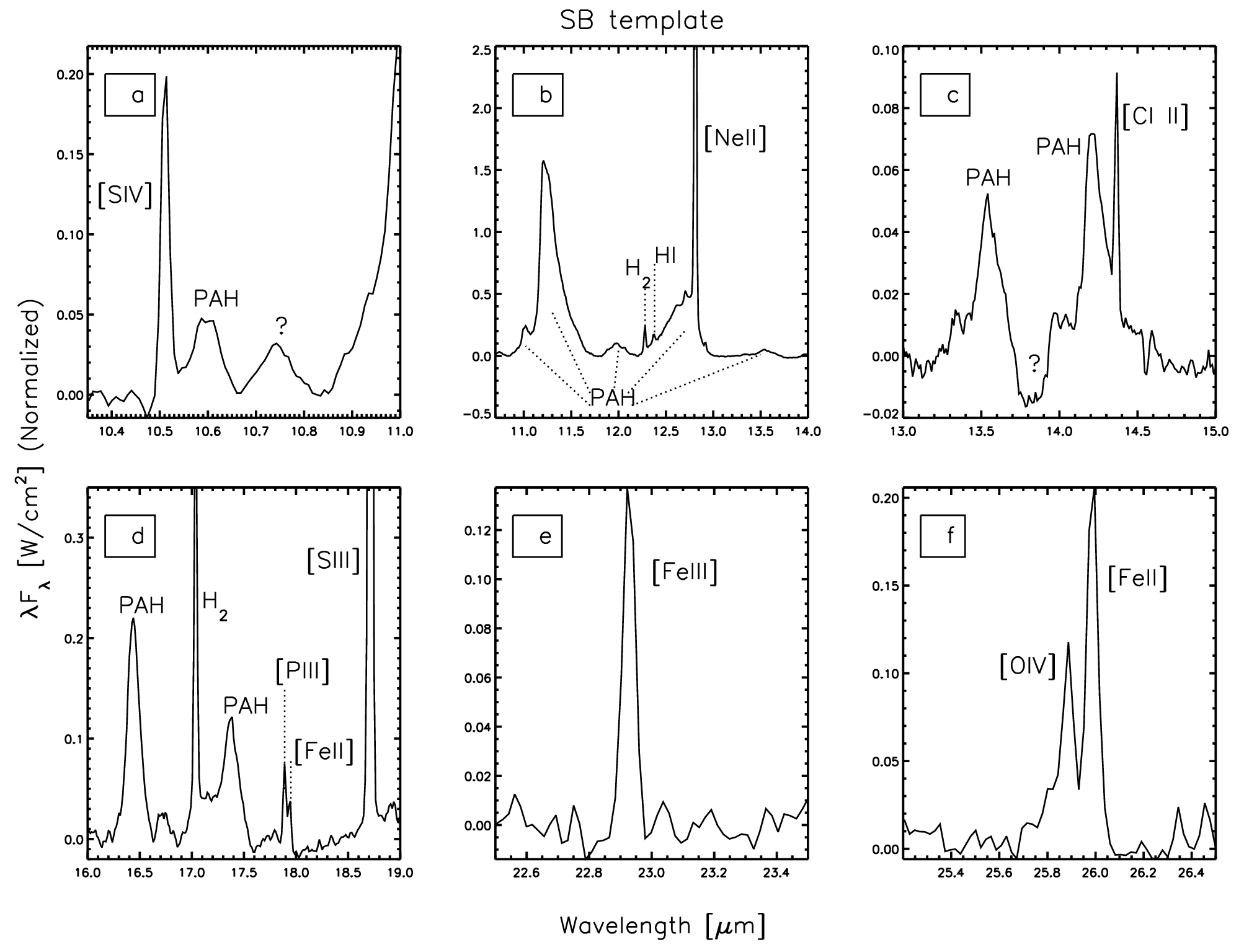}
  \end{center}
  \caption{Close-ups of the continuum subtracted template spectrum.
      The anchor points used to remove the continuum in the individual
      spectra to create this template are those used for the PAH
      integration (see caption of Table~4) plus the following points:
      18.4, 19.4, 21, 22.5, 23.2, 25.0, 26.5, 28.0, 30.0, 32.0, 35.5,
      and 36.5$\mu$m}.
\end{figure*}

\section{Results}\label{results}

  \subsection{Individual Spectra}

  The spectra are shown in Figure~2. The files used to generate this
  plot are available in electronic format (tab1-24.txt files).
  Measurements of fine-structure lines and PAH features that fall
  within the wavelength range of the IRS SH and LH modules are given
  in Tables \ref{lines_hr_s}, \ref{lines_hr_t}, and 4 respectively.

  The spectra contain a variety of emission features. All of the
  spectra contain the prominent fine-structure lines of \ion{[Ne}{2]}
  (12.81$\mu$m), \ion{[Ne}{3]} (15.55$\mu$m), and \ion{[S}{3]} (18.71
  and 33.48$\mu$m). Higher ionization lines such as \ion{[O}{4]}
  (25.98$\mu$m) and \ion{[S}{4]} (10.51$\mu$m) are seen in most of the
  sample, but are weak. The \ion{[Ne}{5]} line at 14.32$\mu$m is
  detected in seven sources.  The \ion{[Ar}{5]} line at 13.1$\mu$m is
  only detected in Mrk\,266. In addition, many objects show emission of
  \ion{[Fe}{2]} (25.99$\mu$m) and \ion{[Fe}{3]} (22.92$\mu$m).  The
  fine-structure lines in the IRS spectra are unresolved (even those
  with AGN), except for Mrk\,266 and NGC\,3079 where there is evidence of
  broadening in the \ion{[Ne}{2]} and \ion{[Ne}{3]} lines (Spoon et
  al. in prep.).  The recombination line HI (6-7) at 12.37$\mu$m is
  detected in 14 objects. Rotational lines of H$_2$, with the
  strongest transition 0-0~S(1) emitting at 17.03$\mu$m, are present
  in most of the spectra. A number of PAH features are seen, the most
  prominent being the neutral C-H out-of-plane bending bands at
  11.2$\mu$m and the 12.7$\mu$m. The 16.4$\mu$m feature and 17$\mu$m
  complex are also detected throughout the sample. The high S/N allow
  us to detect in the spectra other weaker PAH features and a few
  `unidentified' bands. These are best seen in Figures~3 and 4, and
  are discussed in the next section.

  \subsection{Starburst Template}

  We have constructed a high-resolution starburst template spectrum
  from the 15 spectra in our sample with no signs of AGN activity (see
  Table~1). The template is available in electronic format
  (tab25.txt).  This is equivalent to the starburst template derived
  by \citet{Brandl06} but using the IRS high-resolution modules.  To
  create this template, the SH spectrum of each object was first
  scaled up to the LH spectrum using the overlapping continuum around
  19$\mu$m. Based on a spline-defined continuum (with anchor points
  given in caption of Figure~4) each spectrum were separated into a
  continuum-subtracted spectrum (feature spectrum) and the associated
  continuum spectrum (featureless spectrum). The continuum template
  was generated by averaging every individual continuum spectrum after
  normalization to the 14$\mu$m flux.  Likewise, the feature template
  spectrum was created by computing the noise-weighted mean of every
  individual continuum-subtracted spectrum, after normalization to the
  total flux of the emission features.  Finally, the starburst
  template spectrum was assembled by scaling the continuum spectrum to
  the average ratio between the sum of the PAHs and line fluxes over
  the 14$\mu$m continuum before adding it to the feature template
  spectrum. The template is shown in Figure~3.  Up until now, the
  2-40$\mu$m {\em ISO}-SWS spectrum of M82 has been usually adopted in
  the literature as a starburst high-resolution template.  However,
  M82 is a dwarf starburst with a low luminosity
  (3x10$^{10}$~L$_\odot$) while our template is built from a sample of
  15 galaxies spanning a range in IR color and luminosity typical of
  starburst galaxies.  In fact, seven of them are LIRGS
  ($\geq$10$^{11}$~L$_\odot$), which in terms of galaxy evolution are
  much more important as they play a major role in the total energy
  production of the universe at z$\sim$1 \citep{lef05}.  The template
  we created may hence be more representative of the average physical
  parameters of local starburst galaxies, and the average line ratios
  and relative PAH strengths may serve as comparison data for studies
  of starburst galaxies at higher redshifts.
 

  Figure~4 shows a zoom into several interesting parts of the template
  spectrum.  Panel {\em a} shows that, in addition to the \ion{[S}{4]}
  line at 10.51$\mu$m, the spectrum contains the rarely detected
  10.60$\mu$m PAH feature and another broad feature at 10.74$\mu$m.
  Panel {\em b} shows details of the 11.3$\mu$m PAH feature including
  the 11.0$\mu$m component. \citet{slo99} noted that this feature
  (11.0$\mu$m) indicates ionized PAHs based on spatially resolved
  spectra of the reflection nebula NGC\,1333 SVS\,3.  Another PAH
  feature at 12.0$\mu$m is seen followed by the H$_{2}$ 0-0S(2) line
  and the H\,I Hu$\alpha$ recombination line at 12.37$\mu$m.  The
  \ion{[Ne}{2]} line (12.81$\mu$m) sits in the flank of the 12.7$\mu$m
  PAH feature.  In panel {\em c} there is a PAH feature at 13.53$\mu$m
  followed by an unidentified absorption feature (13.65-14.0$\mu$m)
  which has, to our knowledge, not been previously detected. This
  absorption is detected in most of the starbursts but the strongest
  absorptions are found in those sources also harboring an AGN.
    This is shown in Figure~5, where the objects have been ordered
    according to the strength of the absorption feature, and where out
    of the bottom seven objects (those with stronger absorption), six
    have an AGN. Its detection in NGC\,4945 is also confirmed from a
    map study in the region where the dip is seen to change in
    strength at different positions in the central region of the
    galaxy (Spoon et al., in prep). While not reported, this dip is
    also present in the low-resolution starburst template (made from
    the same sample) in \citet{Brandl06}. The weak 14.22$\mu$m PAH
    feature is present and the line at 14.36$\mu$m corresponds to
    \ion{[Cl}{2]}.  Panel {\em d} displays more PAH features at
    16.4$\mu$m, 16.7$\mu$m, and 17.0$\mu$m.  The H$_2$ 0-0~S(1)
    transition at 17.0~$\mu$m is strong and the \ion{[Fe}{2]} line at
    17.98$\mu$m, while weak, is still detected.  Another iron line
    (\ion{[Fe}{3]} at 22.93$\mu$m) is shown in panel {\em e}. The last
    panel shows a blend of lines which correspond to \ion{[O}{4]}
    (25.89$\mu$m) and \ion{[Fe}{2]} (25.98$\mu$m), the stronger of
    which is the iron line.

\begin{figure}\label{weirdabs}
  \begin{center}
  \includegraphics[width=8cm]{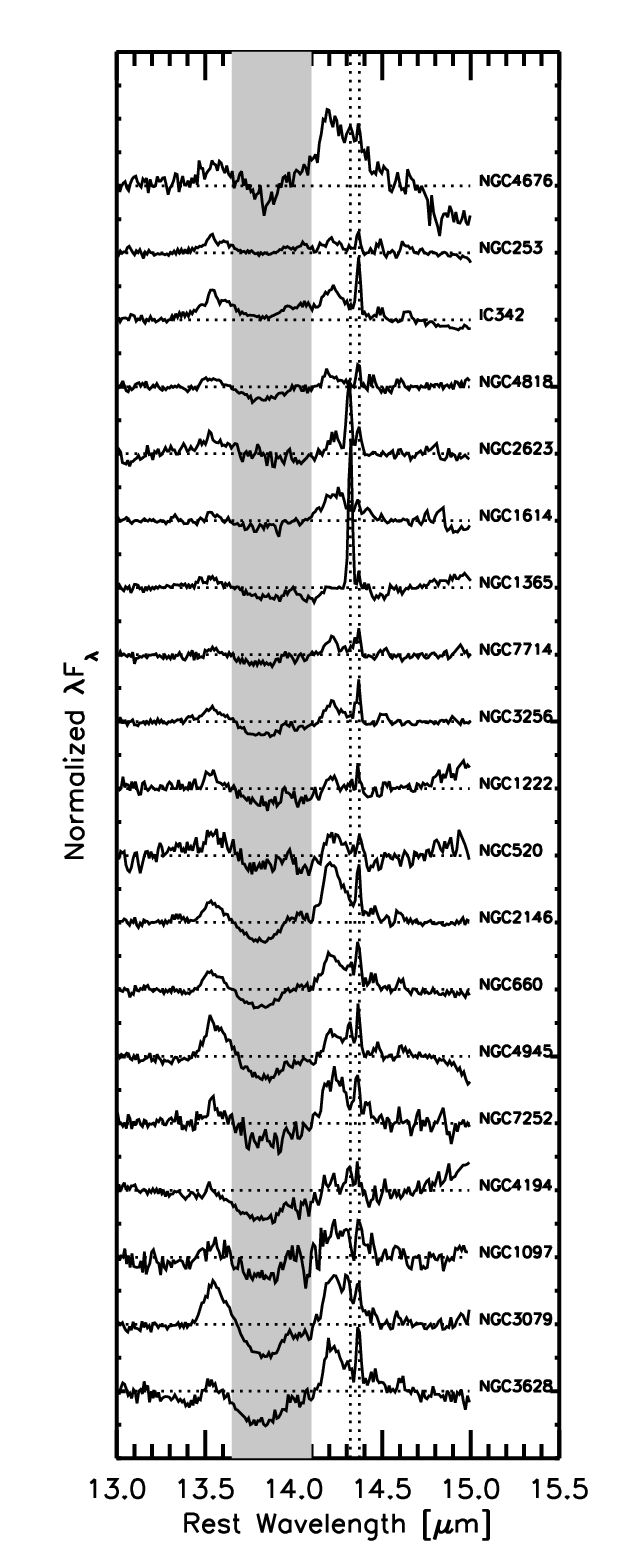}
  \end{center}
  \caption{Continuum divided spectrum to highlight (in grey) the
    absorption feature between 13.56 and 14.0$\mu$m. The sources have
    been ordered by decreasing strength of the feature. The vertical
    lines indicate the position of the \ion{[Ne}{5]} and \ion{[Cl}{2]}
    fine-structure lines.}
\end{figure}

\section{Discussion}\label{discussion}

  \subsection{Conditions in the Ionized Gas}

  The presence of individual lines can set constraints on the
  ionization conditions in the hot gas in our sample. The most notable
  of these is \ion{[Ne}{5]}. The \ion{[Ne}{5]} has a high ionization
  potential (IP) of 97.1~eV and has never been detected in purely
  stellar environments, but is routinely detected in spectra of
  galaxies hosting an AGN
  \citep{sturmetal02,lutzetal03,Weedman05,Dale06} and in about half of
  the 53 ULIRGs analyzed by \citet{Farrah07} using {\em Spitzer} data.
  This line is also seen in young SNe remnants \citep{oli99} but in
  starbursts, where the overall shape of the mid-IR continuum is not
  flat, the presence of [NeV] would imply a weak AGN rather than SNe
  emission (the lifetime of the latter is too small to allow for a
  substantial contribution to the spectrum).  From the objects listed
  in Table~2 previously known to have an AGN component in the
  literature, the \ion{[Ne}{5]} line is present in Mrk\,266,
  NGC\,1365, NGC\,2623, NGC\,3079, NGC\,3628, and NGC\,4945 \citep[see
  also][]{Dudik07}, but not in NGC\,660 and NGC\,1097. In these two
  galaxies the \ion{[Ne}{5]}14.3$\mu$m/\ion{[Ne}{2]}12.8$\mu$m line
  ratio is $<$0.07 hinting at an AGN contribution of less than a few
  percent using the diagram by \citet{Farrah07} (their Fig.~16).
  NGC\,1097 is a particularly interesting case; it also does not show
  \ion{[O}{4]} or \ion{[S}{4]}, hence the AGN is likely very weak in
  the mid-IR. It either has a complex extinction pattern around it or
  it is overwhelmed by the starburst.  We further detect \ion{[Ne}{5]}
  in NGC\,4194, which has no previous AGN classification.

\begin{figure}
  \begin{center}
  \includegraphics[width=6.5cm,angle=90]{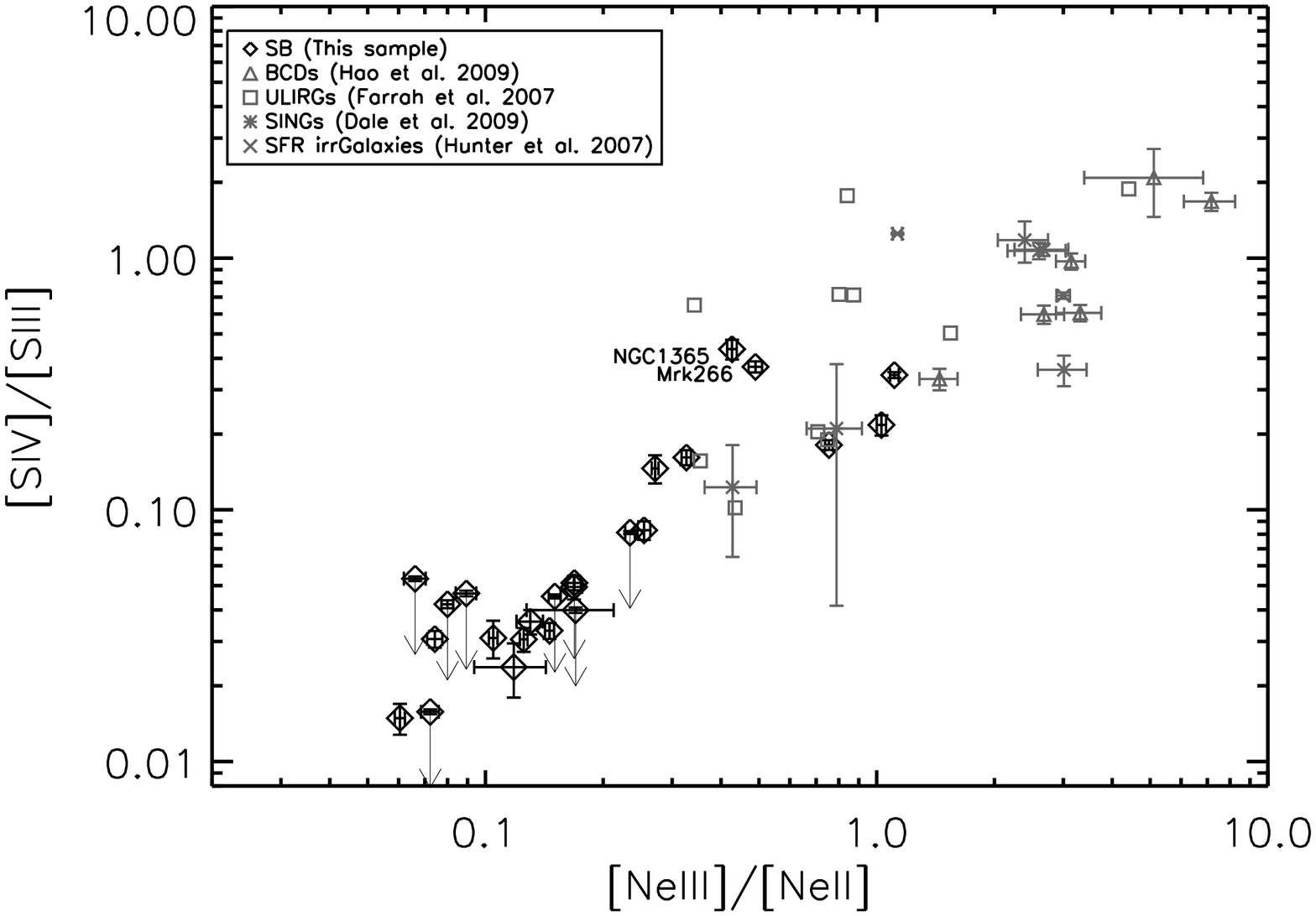}
  \end{center}
  \caption{Hardness of the radiation field as traced by the sulfur
    (10.51$\mu$m/18.71$\mu$m) and neon (15.55$\mu$m/12.81$\mu$m) line
    ratios.}
\end{figure}

\begin{figure}
  \begin{center}
  \includegraphics[width=6.5cm,angle=90]{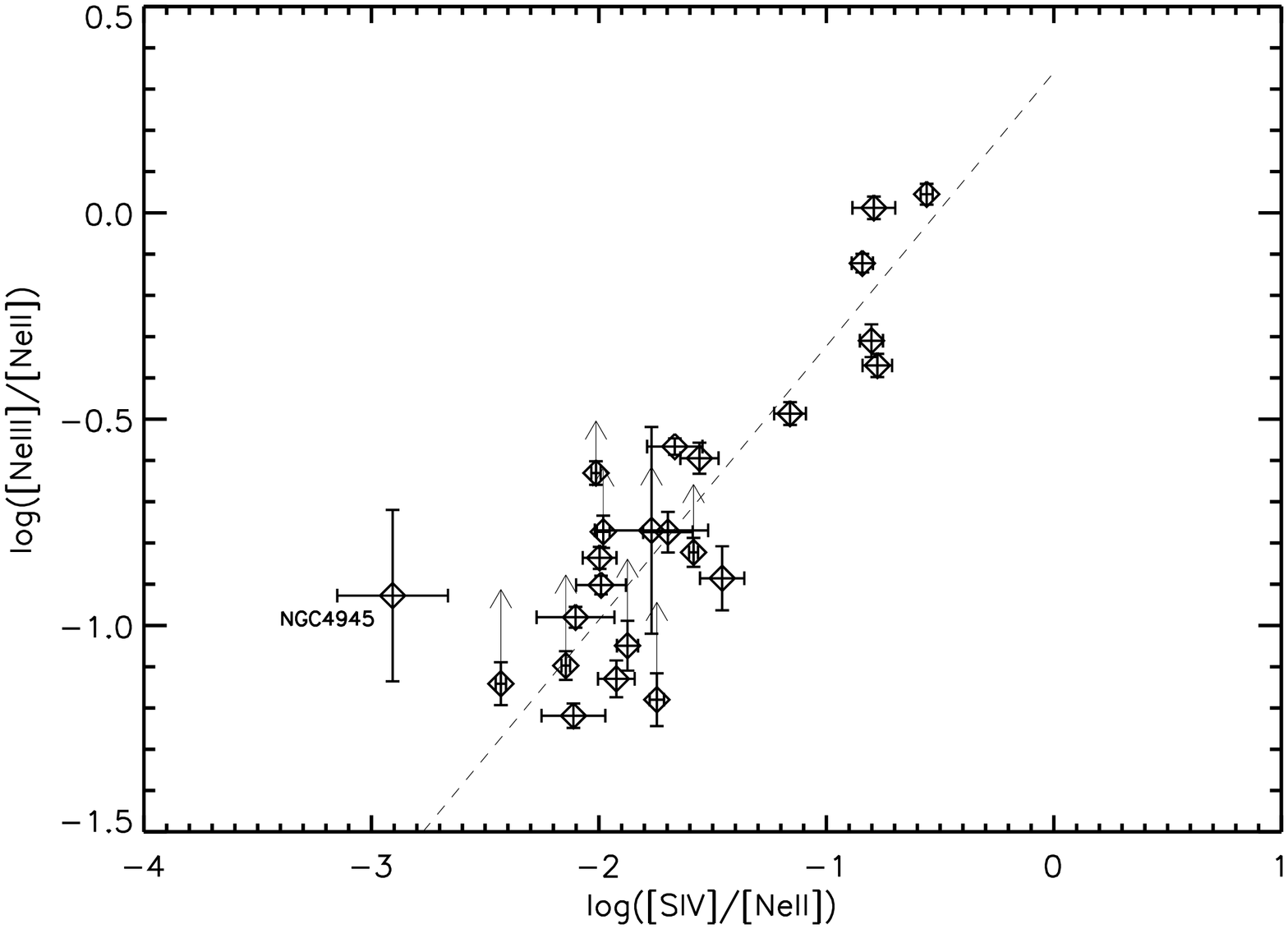}
  \end{center}
  \caption{Hardness of the radiation field as given by the
    \ion{[Ne}{3]}/\ion{[Ne}{2]} and \ion{[S}{4]}/\ion{[Ne}{2]} ratios.
    The dashed line indicates the best fit to the data excluding NGC\,4945:
    $\log([NeIII]/[NeII])=-0.38(\pm0.12)+0.68(\pm0.08)\times\log([SIV]/[NeII])$.}
\end{figure}

\begin{figure}
  \begin{center}
  \includegraphics[width=6.5cm,angle=90]{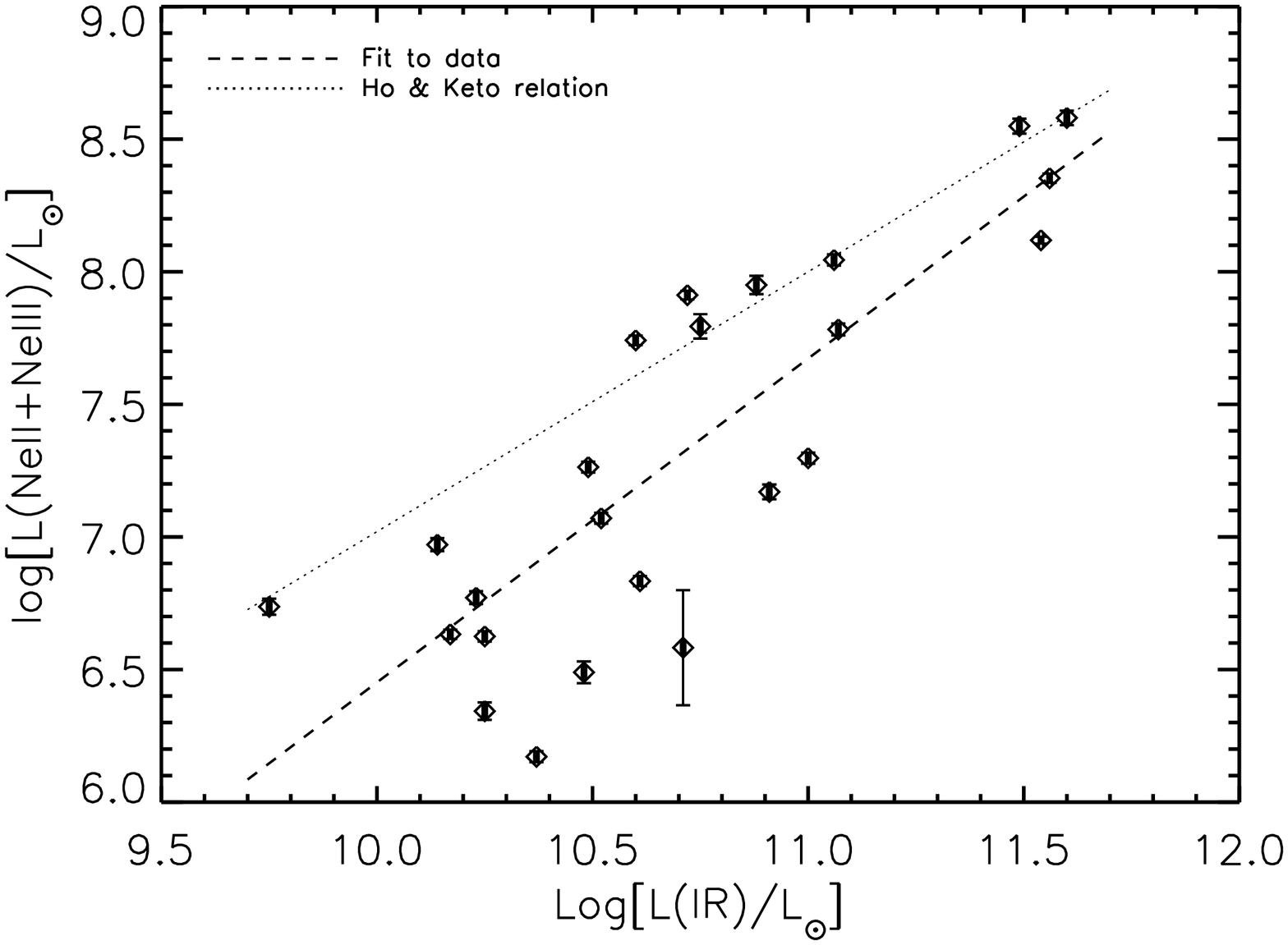}
  \end{center}
  \caption{Sum of the \ion{[Ne}{2]} and \ion{[Ne}{3]} luminosities
    plotted against the infrared luminosity (8-1000~$\mu$m). The
    dashed line indicates the best fit to the data:
    $\log[L(NeII+NeIII)/L_\odot]=-5.3(\pm2.1)+1.18(\pm0.20)\times\log[L(IR)/L_\odot]$.}
\end{figure}

  The strongest detections of \ion{[Ne}{5]} relative to \ion{[Ne}{2]}
  in our sample are found in Mrk\,266 and NGC\,1365, at ratios of
  0.140$\pm$0.005 and 0.135$\pm$0.005, roughly a factor 10 above the
  rest of the sample.  Adopting the AGN-starburst mixing scenarios of
  Sturm et al. (2002) and Farrah et al. (2007), this would imply the
  AGN contribution to be 10 times larger in these two sources compared
  to the rest of the sample.


  Two other individual lines are worth mentioning; the \ion{[S}{4]}
  line, and the 18$\mu$m and 26$\mu$m \ion{[Fe}{2]} lines. The
  \ion{[S}{4]} line is present in most of the objects but is always
  weak.  The line has an IP of 34.8~eV, below that of the
  \ion{[Ne}{3]} line, and therefore should be easily excited. However,
  the photo-ionization cross-section of S$^{3+}$ is lower than
  Ne$^{2+}$, and one needs many photons to produce the \ion{[S}{4]}
  line. The IP of \ion{Fe}{2} is even lower, at 7.87~eV.  It could
  therefore ionize within the photo-dissociation region (PDR).
  \citet{leb08} have shown that this ion correlates well with the
  \ion{[Ar}{2]} line which is a good tracer of the ionized gas near
  the PDR.

\begin{figure}{!ht}
  \begin{center}
  \includegraphics[width=8.5cm]{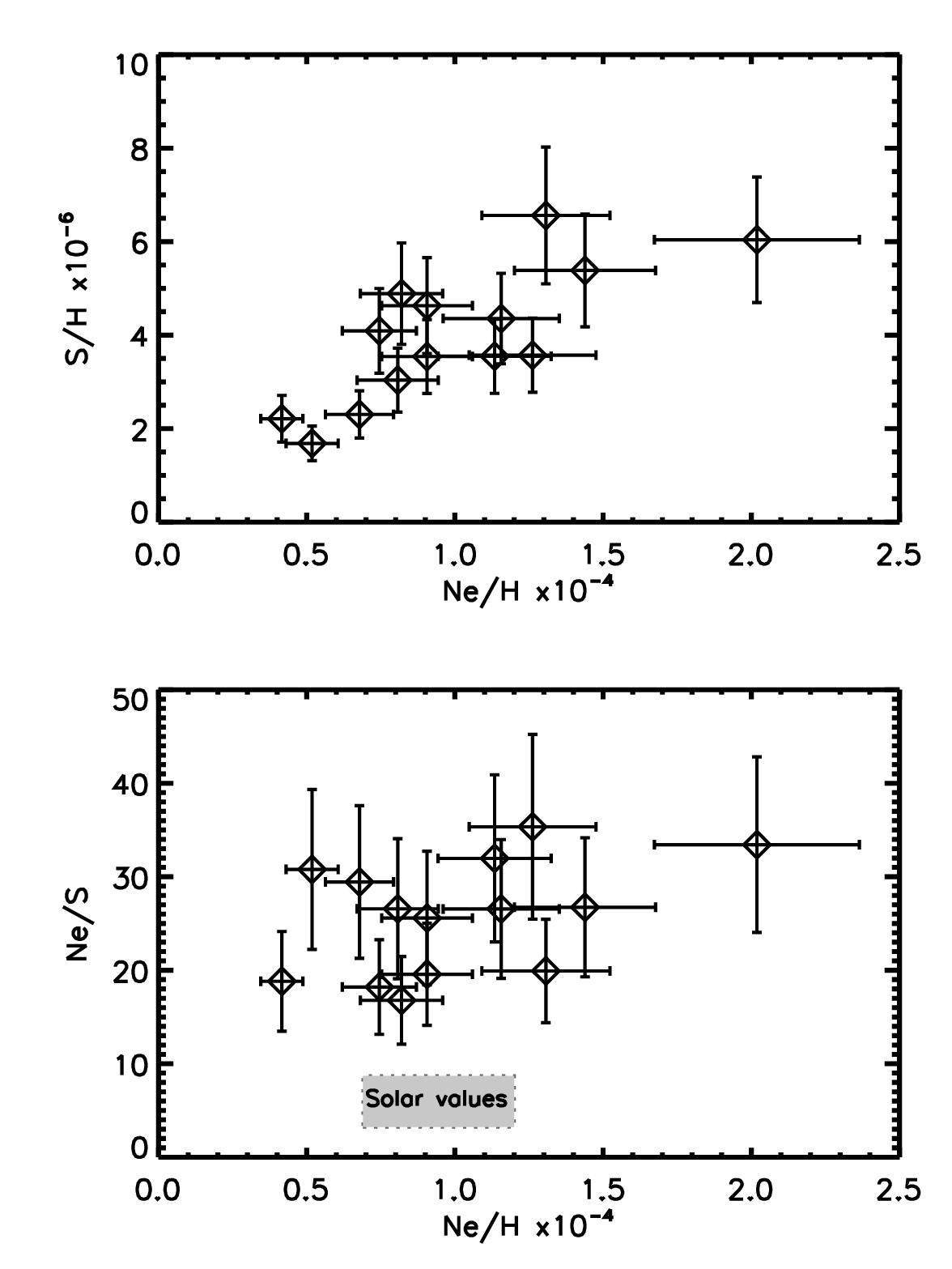}
  \end{center}
  \caption{ Abundance relations for those objects with a measured
      HI line in the IRS spectrum. These include 14 galaxies of which three have an AGN.
    Top panel: Neon and sulfur abundances. Bottom panel: Neon over
    sulfur ratio plotted against the neon abundance as an indicator of
    the metallicity.  The grey box in the bottom panel indicates
      the range of solar values found in the literature (see \S5.2).
      Note that the solar sulfur abundance ranges from 1.4 to
      2.1$\times$10$^{-5}$, which is significantly higher that the range of
      values in the starburst sample (top panel) and is therefore not
      plotted.}
\end{figure}

\begin{figure}{!t}
  \begin{center}
  \includegraphics[width=6.5cm,angle=90]{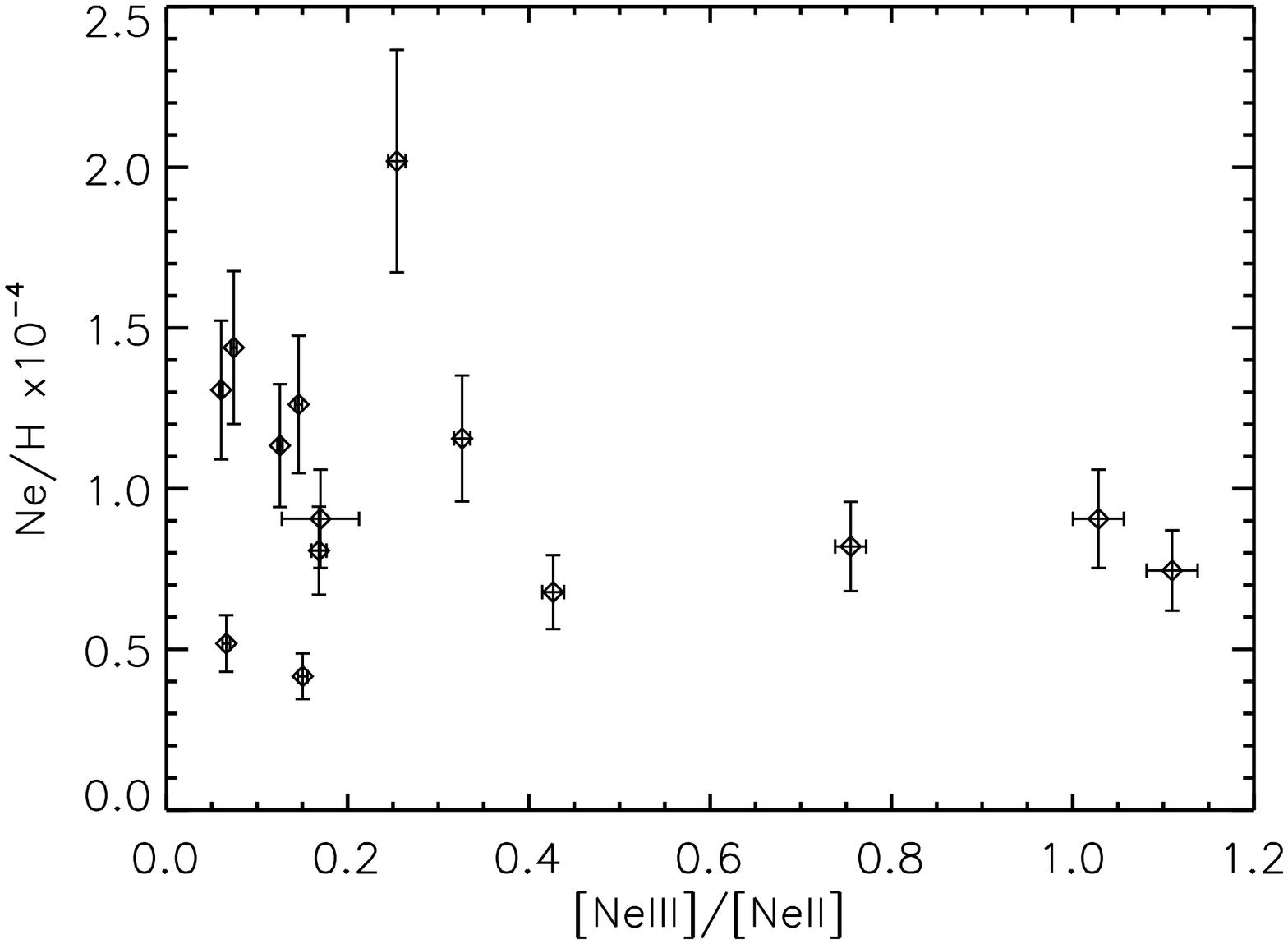}
  \end{center}
  \caption{Neon abundance versus the hardness of the radiation field.}
\end{figure}

\begin{deluxetable*}{l c c c c c c c c c c}{!h}
 \tabletypesize{\scriptsize}
   \tablewidth{0pt}
   \tablecaption{Elemental Abundances\tablenotemark{a}.\label{lines_abun_t}}
   \tablehead{\colhead{Object} & \colhead{S$^{2+}$/H}  &
     \colhead{S$^{3+}$/H}  & \colhead{Ne$^{+}$/H}  &
     \colhead{Ne$^{++}$/H}  &  \colhead{Fe$^{+}$/H}  &  
     \colhead{Fe$^{++}$/H} & ICF  & 
     \colhead{S/H\tablenotemark{b}}  & \colhead{Ne/H}  & \colhead{Fe/H}\\
     & \colhead{(10$^{-6}$)} & \colhead{(10$^{-7}$)} & \colhead{(10$^{-4}$)}  &
     \colhead{(10$^{-5}$)} & \colhead{(10$^{-6}$)} &
     \colhead{(10$^{-6}$)} & 
     \ion{Fe}{4} & \colhead{(10$^{-6}$)} & \colhead{(10$^{-4}$)} & \colhead{(10$^{-6}$)}}
   \startdata

NGC\,1097 &   3.22 &$<$0.22 &   0.84 &   0.68 &   1.18 &   0.56 &   1.09 &   3.54 &   0.91 &   1.90 \\
        &   0.79 &   0.00 &   0.14 &   0.11 &   0.19 &   0.16 &        &   0.79 &   0.15 &   0.35 \\
NGC\,1222 &   3.53 &   2.09 &   0.49 &   2.58 &   0.40 &   0.78 &   1.58 &   4.09 &   0.75 &   1.87 \\
        &   0.87 &   0.37 &   0.08 &   0.43 &   0.07 &   0.22 &        &   0.91 &   0.12 &   0.29 \\
NGC\,1365 &   1.96 &   1.47 &   0.56 &   1.15 &   0.56 &   0.31 &   1.17 &   2.30 &   0.68 &   1.02 \\
        &   0.48 &   0.25 &   0.10 &   0.19 &   0.09 &   0.08 &        &   0.50 &   0.12 &   0.17 \\
  IC\,342 &   5.95 &   0.15 &   1.27 &   0.37 &   1.47 &   1.34 &   1.07 &   6.56 &   1.31 &   3.01 \\
        &   1.46 &   0.03 &   0.21 &   0.06 &   0.24 &   0.38 &        &   1.46 &   0.22 &   0.62 \\
NGC\,1614 &   5.42 &   0.78 &   1.80 &   2.19 &   2.23 &   2.94 &   1.20 &   6.04 &   2.02 &   6.21 \\
        &   1.33 &   0.14 &   0.31 &   0.36 &   0.36 &   0.82 &        &   1.34 &   0.35 &   1.18 \\
NGC\,2146 &   3.23 &   0.18 &   1.18 &   0.82 &   0.64 &   0.37 &   1.09 &   3.57 &   1.26 &   1.10 \\
        &   0.79 &   0.03 &   0.20 &   0.14 &   0.10 &   0.10 &        &   0.79 &   0.21 &   0.21 \\
NGC\,3256 &   3.21 &   0.17 &   1.07 &   0.64 &   0.71 &   0.87 &   1.13 &   3.55 &   1.13 &   1.78 \\
        &   0.79 &   0.03 &   0.18 &   0.11 &   0.12 &   0.25 &        &   0.79 &   0.19 &   0.37 \\
NGC\,3310 &   4.07 &   1.53 &   0.61 &   2.98 &   1.08 &   0.90 &   1.38 &   4.63 &   0.91 &   2.73 \\
        &   1.00 &   0.27 &   0.10 &   0.50 &   0.18 &   0.25 &        &   1.03 &   0.15 &   0.43 \\
NGC\,3556 &   2.01 &$<$0.16 &   0.39 &   0.28 &   0.21 &   0.19 &   1.13 &   2.21 &   0.42 &   0.45 \\
        &   0.50 &   0.00 &   0.07 &   0.05 &   0.04 &   0.06 &        &   0.50 &   0.07 &   0.09 \\
NGC\,4088 &   1.53 &$<$0.14 &   0.50 &   0.16 &   0.30 &   0.18 &   1.06 &   1.68 &   0.52 &   0.51 \\
        &   0.37 &   0.00 &   0.09 &   0.03 &   0.05 &   0.05 &        &   0.37 &   0.09 &   0.10 \\
NGC\,4194 &   3.86 &   1.08 &   1.00 &   1.56 &   0.69 &   1.06 &   1.25 &   4.35 &   1.16 &   2.19 \\
        &   0.95 &   0.18 &   0.17 &   0.26 &   0.11 &   0.30 &        &   0.97 &   0.20 &   0.41 \\
NGC\,4676 &   2.74 &   0.23 &   0.75 &   0.60 &   0.65 &   0.00 &   1.00 &   3.04 &   0.81 &   0.65 \\
        &   0.68 &   0.04 &   0.13 &   0.10 &   0.11 &   0.00 &        &   0.68 &   0.14 &   0.11 \\
NGC\,4818 &   4.87 &   0.26 &   1.39 &   0.49 &   2.09 &   1.99 &   1.08 &   5.38 &   1.44 &   4.43 \\
        &   1.20 &   0.04 &   0.23 &   0.08 &   0.35 &   0.58 &        &   1.21 &   0.24 &   0.92 \\
NGC\,7714 &   4.32 &   1.35 &   0.60 &   2.17 &   0.71 &   1.22 &   1.43 &   4.89 &   0.82 &   2.76 \\
        &   1.06 &   0.24 &   0.10 &   0.37 &   0.12 &   0.34 &        &   1.08 &   0.14 &   0.46 \\
 Solar\tablenotemark{c}   &   .... &   .... &   .... &   .... &   .... &   .... &   .... &   1.4-2.1 &   0.7-1.2  &  32.4 \\

  \enddata

  \tablenotetext{a}{Abundances are given in number with the
    uncertainties indicated in the row below each measurement.}

  \tablenotetext{b}{Includes 10\% of \ion{S}{3} to account for the
    contribution of \ion{S}{2} (see \S5.2).}

  \tablenotetext{c}{Range of solar abundances taken from
    \citet{gre98,fel03,asp05,asp06}. See \S5.2 for explanation.}

\end{deluxetable*}

  Fine structure lines of the same species but from different
  ionization stages can be used to trace the properties (hardness and
  ionization parameter) of the ionizing radiation field.  The diamond
  symbols in Figure~6 shows the ratio of the \ion{[S}{4]}10.5$\mu$m
  (IP$=$34.8~eV) and \ion{[S}{3]}18.7$\mu$m (IP$=$23.3~eV) lines
  plotted against the ratio of the \ion{[Ne}{3]}15.5$\mu$m
  (IP$=$41.0~eV) and \ion{[Ne}{2]}12.8$\mu$m (IP$=$21.6~eV) lines for
  the starburst sample.  As expected both ratios trace each other.
  Exceptions are Mrk\,266 and NGC\,1365.  These two outliers both have
  a strong AGN contribution to the neon lines, as measured by their
  $>$10 times elevated \ion{[Ne}{5]}/\ion{[Ne}{2]} ratio compared to
  the other sources.  Excluding these two sources, the slope for the
  correlation between \ion{[S}{4]}/\ion{[S}{3]} and
  \ion{[Ne}{3}/\ion{[Ne}{2]} is 0.25$\pm$0.03. Our values values fall
  within the range of values found by \citet{vermaetal03} in their
  {\em ISO} study of starburst galaxies (including NGC\,1365).
  Figure~6 also includes representative subsamples of Blue Compact
  Dwarf Galaxies \citep[BCDs][]{hao09}, ULIRGs \citep{Farrah07}, and
  galaxies from the Spitzer Infrared Nearby Galaxy Survey
  \citep[SINGS][]{dal09}.  The ratios in the SINGS galaxies studied by
  \citet{dal09} and in several star forming regions in irregular
  galaxies analyzed by \citet{hun07} follow the trend of the starburst
  sample.  The ratios in the starburst galaxies are lower than the BCD
  sample studied by \citet{wu08}, and in fact Figure~6 shows only the
  lowest BCD ratios from their sample.  \citet{hao09} compare in
  detail our starburst sample with similarly analyzed samples of
  ULIRGs, BCDs, and AGNs. They find that the starburst galaxies have
  the lowest excitation and are located in the lower branch of this
  diagram.  Compared to the Starburst and ULIRG samples of
  \citet{vermaetal03} and \citet{Farrah07} the excitation level in our
  sample is on the low and middle end, with the highest ratios being
  \ion{[S}{4]}/\ion{[S}{3]}$\sim$0.4 and
  \ion{[Ne}{3]}/\ion{[Ne}{2]}$\sim$1, and the \citet{vermaetal03} and
  \citet{Farrah07} samples reaching values as high as 10 for both
  ratios. This lower ionization in our sample may indicate that there
  are fewer massive stars or that they are older.  Interestingly, the
  three sources with the highest \ion{[Ne}{3]}/\ion{[Ne}{2]} ratios in
  our sample ($>$0.6) are not those with an AGN component, but pure
  starbursts.  NGC\,1222, NGC\,3310 and NGC\,7714 all are found on the
  `BCD branch' of the \ion{[Ne}{5]}/\ion{[Ne}{2]} versus
  \ion{[Ne}{3]}/\ion{[Ne}{2]} diagnostic diagram of \citet{hao09}.

  \citet{gro08} studied the relation between the well known
  mid-infrared ionization diagnostic with the ground accessible ratio
  of \ion{[S}{4]}/\ion{[Ne}{2]}. They found a good correlation between
  the ratios in their sample, which included starburst galaxies, BCDs,
  ULIRGs, AGNs, Planetary Nebulae, and Galactic and extra-galactic
  H\,II regions. We confirm this relation (Figure~7), and find a slope
  value of 0.68$\pm$0.08 (excluding NGC\,4945), in good agreement with
  the slope found by \citet{gro08} (0.65 for the starburst sub-sample
  and 0.81 for their full sample).  NGC\,4945 has the strongest silicate
  absorption feature of the sample \citep{Brandl06} and this
  attenuates the \ion{[S}{4]} line flux relative to the \ion{[Ne}{2]}
  line which is exactly what is seen in Figure~7.

  \citet{ho07} illustrated how the neon lines in the mid-infrared
  spectra can be a powerful indicator to derive the star formation
  rate in star-forming galaxies. They showed that the sum of the
  luminosities of both lines correlate well with the infrared
  luminosity in over 5 orders of magnitude in luminosity, and proposed
  a calibration between the luminosity of the lines and the star
  formation rate.  Figure~8 shows the relation between the neon lines
  and infrared luminosity. Our data covers a narrow range of infrared
  luminosities, but it can be seen that a correlation is present. The
  slope in the figure has a value of 1.18$\pm$0.20, agreeing within
  the errors with that reported by \citet{ho07} (0.98$\pm$0.069).

  \subsection{Abundances}

  Infrared lines offer many advantages over optical or ultraviolet
  (UV) lines for measuring elemental abundances, the most important of
  which is their small dependence on the electron temperature
  \citep[e.g.][]{ber01,wu08}. This reduces the uncertainty in the
  abundance estimates when there are electron temperature variations,
  which is the case in integrated spectra of galaxies, or when the
  electron temperature cannot be measured. IR lines are also less
  affected by extinction. Optical and UV studies also rely on the
  often rather uncertain, ionization correction factors (ICFs) to
  account for unobserved stages of ionization. The direct measure of
  important stages of ionization of some elements in the infrared
  greatly reduces, or completely avoids, the use of ICFs.


  The HI (6-7) line at 12.37$\mu$m is detected in 14 of the stabursts.
  This allows us to measure the abundances of neon, sulfur, and
  (tentatively) iron relative to hydrogen. The abundances are derived
  adopting an electron temperature (T$_e$) of 10\,000~K and a 
    gas density of 100~cm$^{-3}$. These are values typical of H\,II
  regions.  We derived abundances for a range of T$_e$
  (7500-12\,500~K), and used the differences to obtain the
  uncertainties in the measurements.  These uncertainties amount to
  around 20\% in the total abundances.  Using these parameters and
  solving the equation of statistical equilibrium for the levels of
  the atoms, the ionic abundance can be derived \citep[see
  eq.~1][]{ber01}. These values are reported in
  Table~\ref{lines_abun_t}. Note that we cannot derive the density
  from the pair of \ion{[S}{3]} (18.7$\mu$m/33.4$\mu$m) or
  \ion{[Ne}{3]} (15.5$\mu$m/36.0$\mu$m) lines because the dependence
  of these ratios on the density starts above 3000~cm$^{-3}$ for the
  sulfur ratio, and higher for the neon ratio.
    Compared to our reported values, an electron density of
    1000~cm$^{-3}$ (on the high side for H\,II regions) will produce
    abundances that are no more than 1.7\% higher for Neon, 13\%
    higher for Sulfur, and about 80\% lower for Iron.  The latter is
    due to the electronic configuration of iron.

\begin{figure*}{!h}
  \begin{center}
  \includegraphics[width=16cm]{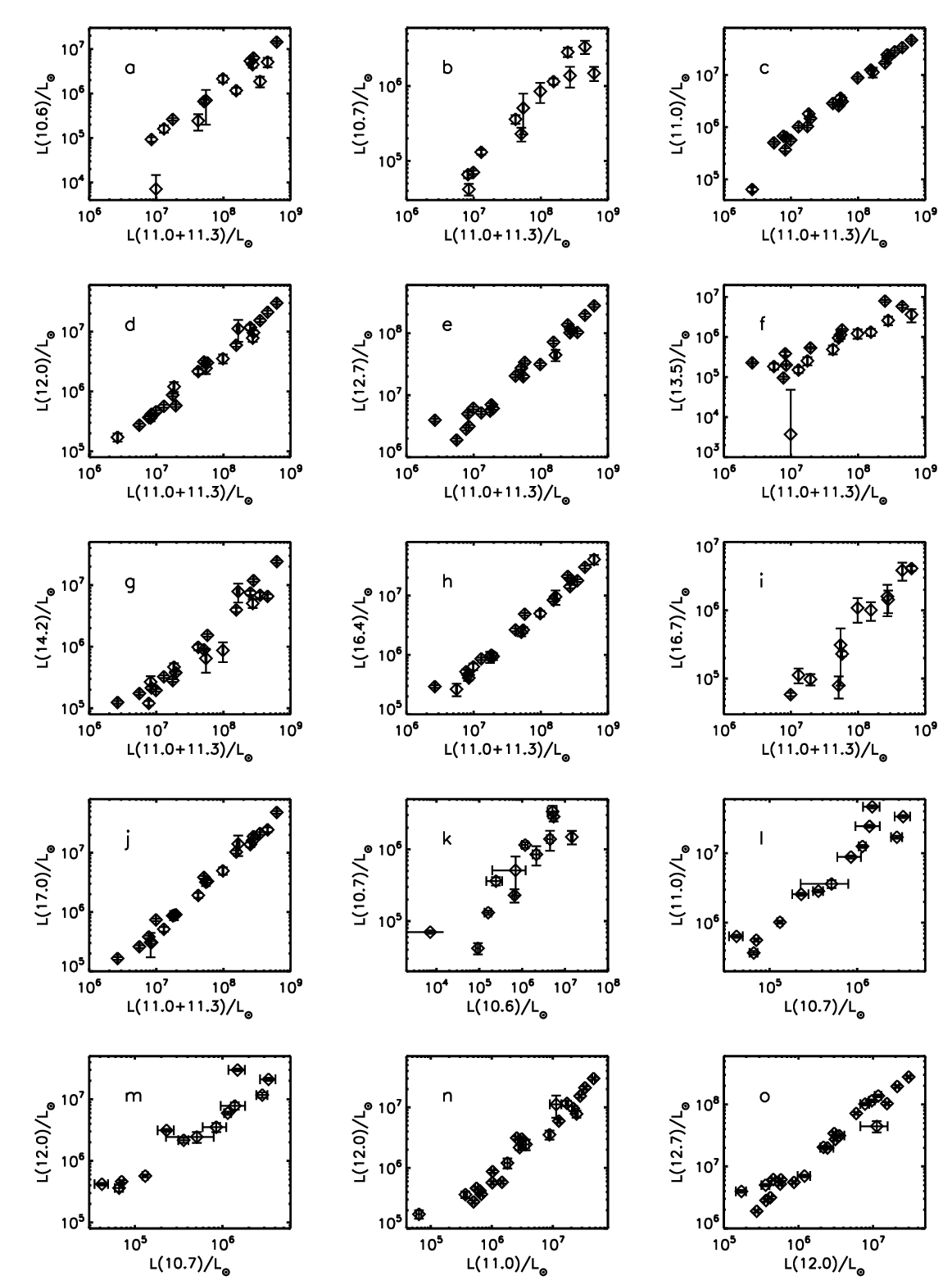}
  \end{center}
  \caption{Correlation between the PAH bands.}
\end{figure*}

    In starburst galaxies the most important stages of ionization of
    neon for abundance determination are Ne$^{+}$ and Ne$^{++}$ which
    are measured in the IR.  Thus, no ICFs are needed. Among the 14
    objects with a HI detection, Ne$^{4+}$ is present in NGC\,1365 and
    NGC\,4194 (Table~\ref{lines_abun_t}), but this likely originates
    from the AGN given the hard radiation field needed to break the
    high ionization potential (IP) barrier of 97.1~eV to produce this
    ion.  This applies as well to the Ne$^{3+}$ (IP of 63.5~eV, also
    above the double ionized helium cutoff). In addition, the
    contribution of Ne$^{4+}$ to the total neon abundance is less than
    1\%. The \ion{[Ne}{5]}14.3$\mu$m/\ion{[Ne}{2]}12.81$\mu$m ratio in
    NGC\,1365 and NGC\,4194 is 0.13 and 0.018 respectively. These
    ratios correspond in the \citet{Farrah07,Spoon09} diagrams to an
    AGN contribution of $\sim$10\% and $<$1\% respectively. This
    implies that the abundances we derive for NGC\,4194 is indeed
    representative of the starburst while in NGC\,1365 our abundances
    may be $\sim$10\% too high. The total sulfur abundance is derived
    adding the contribution from S$^{++}$ and S$^{3+}$. No S$^{+}$ is
    measured in the infrared but its contribution to the total sulfur
    abundance is usually small.  \citet{mar02} estimated a
    contribution of about 15\% of S$^{+}$ in the Galactic
    ultra-compact HII regions they studied.  \citet{ver02} analyzed a
    sample of Magellanic Cloud H\,II regions using IR data, and
    combined optical measurements of \ion{[S}{2]} lines. They derived
    an ionic abundance of S$^{+}$ between 2 to 20\% of the total
    sulfur abundance, with $\sim$8\% being the average.  Using the
    photo-ionization code of \citet{sta90} we have created a grid of
    models at typical conditions of H\,II regions; stellar
    temperatures considered between 37\,500 to 55\,000~K, and electron
    density of 10~cm$^{-3}$, and a filling factor ranging from 0.2 to
    1.0. These models predict the contribution of S$^{+}$ to be
    between 7 and 15\% of that of S$^{++}$ (which is the dominant
    contributor to the total sulfur abundance). We have thus adopted a
    10\% contribution of S$^{+}$ to the sulfur abundance (ICF$=$1.1)
    to account for the contribution of this ion in our sample.  While
    we have measured Fe$^{+}$ and Fe$^{++}$ from our {\em Spitzer}
    data, Fe$^{3+}$ may be an important contributor (up to half) to
    the total iron.  Its contribution has been estimated using the
    photo-ionization grid models by \citet{sta90}, and taking
    advantage of the similarity in IP between the
    \ion{[Ne}{3]}/\ion{[Ne}{2]} and \ion{[Fe}{4]}/\ion{[Fe}{3]} line
    ratios to scale the contribution of this ion in each object. At
    can be seen from Table~5 the ICF varies from a few percent to a
    factor 2. Given the importance of this ion to the total iron
    budget in those objects with a high ICF, the iron abundances we
    report should be interpreted with care.  The uncertainty for this
    element in Table~5 only takes into account the uncertainty in the
    Fe$^{+}$ and Fe$^{++}$ ionic abundance.

    In Figure~9 (top), the neon abundance of our starburst sample
    correlates reasonably well with the sulfur abundance.  This is
    expected because they are both $\alpha$-elements and are being
    produced in the same type of stars.  This finding was also pointed
    out by \citet{wu08} in their BCD sample and it contradicts earlier
    findings \citet{vermaetal03} who based on {\em ISO} data did not
    find such a correlation between the abundance of these two
    elements for the starbursts they studied.
    In our sample the neon abundances range from slightly supersolar
    (IC\,342, NGC\,1614, NGC\,4818) to several times lower than the
    solar value (see Table~5).  Note that the solar abundances of
    certain elements have been subject to many changes during the last
    decade \citep{pot06, ber08, leb08}, and thus, we resort to compare
    to the range of values found in the literature
    (0.69-1.2$\times$10$^{-4}$).  There is however increasing evidence
    that the higher solar neon abundance reported agrees better with
    the findings of studies in Galactic H\,II regions and planetary
    nebulae. The sulfur abundance is also lower than solar (more so
    than neon, with solar sulfur values ranging from
    1.4-2.1$\times$10$^{-5}$), which is in agreement with many
    previous studies.  Available optical-derived abundances of oxygen
    by \citet{pil04,pas93,gon95} in IC\,342, NGC\,3310, and NGC\,7714
    (7.1, 1.5,1.8$\times$10$^{-4}$ respectively) confirm the trend of
    our IR-derived neon abundances, with IC\,342 supersolar, and the
    other two galaxies having a 2-3 times lower abundance than the Sun
    (adopting 4.9$\times$10$^{-4}$ for the solar oxygen abundance).

    The bottom panel in Figure~9 displays the Ne/S ratio versus the
    neon abundance. The range of solar values is indicated by the grey
    box in the figure. The Ne/S ratio varies from 18 to 40. This range
    of values is typical for other infrared derived abundances in
    starburst galaxies, H\,II regions and even planetary nebulae
    \citep{rub07,rub08,ber08}. It is overall higher than the ratio
    found by \citet{wu08} in their abundance analysis of BCDs, where
    they find values from 10-20. The Ne/S ratios are higher than the
    solar Ne/S ratio, revealing a discrepancy between the neon and
    sulfur solar values, and which has now been widely reported
    \citep[e.g.][and references
    therein]{mar03,hen04,rub07,ber08,wan08}.  In Figure~10 the neon
    abundance is plotted versus the hardness of the radiation field.
    The sample is small but seems to indicate that a hard radiation
    field is found only for sources with a low metallicity which ties
    well with the common knowledge that at low stellar metallicity,
    line blanketing is diminished resulting  in harder stellar
    radiation fields.

  \subsection{PAHs}

  In Figure~11 the PAH luminosity for the different bands is plotted.
  We chose to compare the PAH features in the spectra to the
    integrated emission of the 11.0$+$11.3$\mu$m complex because it is
    the strongest in this wavelength region (panels `a' through `j').
    Panels e, h, and j show a very good correlation of the 11.3$\mu$m
    complex with the relatively strong PAH bands at 12.7, 16.4, and
    the 17.0$\mu$m PAH complex. The 17.0$\mu$m complex was discovered
    by \citet{ker00} as a weak feature in {\em ISO} spectra and
    attributed to PAHs by laboratory analysis \citep{Leger84, all85}.
    \citet{Smith07} corroborate this using spectra from the SINGS
    galaxies and showing good correlation between the strength of the
    17.0$\mu$m and the 11.3$\mu$m PAH complexes. Our sample supports
    this finding (panel `j').  Other bands also correlate well (e.g
    11.0, 12.0, 16.7$\mu$m). The correlation of the weaker bands
    (10.6, 13.5, 14.2$\mu$m) is good but shows more scatter.  The
    16.7$\mu$m feature is listed by \citet{Smith07} as part of the
    17.0$\mu$m complex but in our higher resolution spectra we can
    separate it from this complex.  \citet{Smith07} also report a PAH
    feature at 15.9$\mu$m.  This feature is not present in our high
    S/N template spectrum but may be present in NGC\,3079, NGC\,3628
    and NGC\,4676, albeit weakly.  The unidentified feature at
    10.75$\mu$m (panel `b') shows a good correlation with the
    11.3$\mu$m complex.  In the bottom panels (`l' and `m') we plot
    this feature with respect to other weak PAHs. Although there is
    some scatter, they do seem to correlate.  It is therefore likely
    that the 10.75$\mu$m feature is either a PAH or at least its
    carrier is carbonaceous in nature.  Panel `n' shows a very good
    correlation between the satellite features in the 11.0 and
    12.0$\mu$m bands, and the last panel displays the equally good
    correlation between the 12$\mu$m PAHs. In summary, the strengths
    of the PAH features scale with each other.

\begin{figure}
  \begin{center}
  \includegraphics[width=9cm]{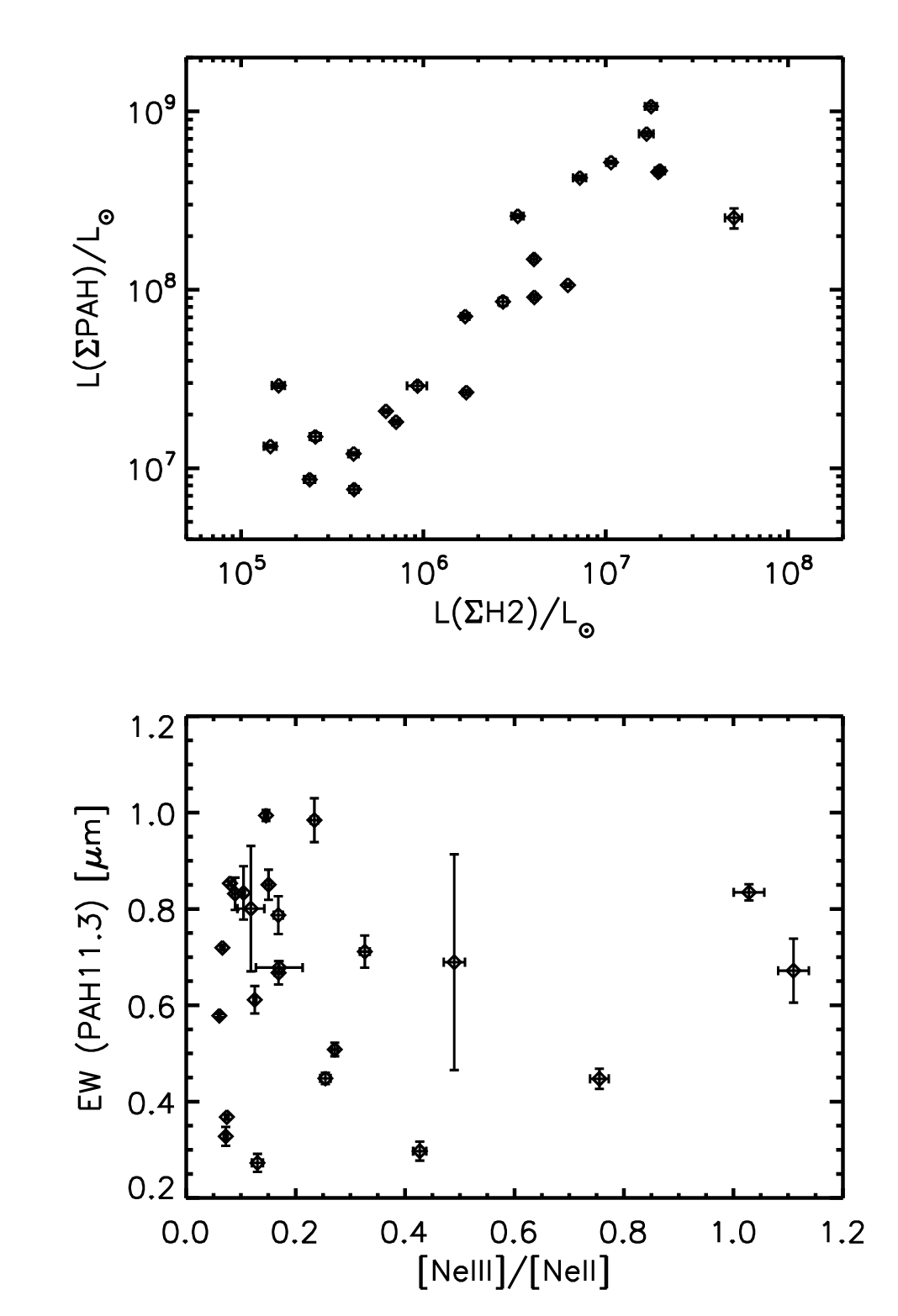}
  \end{center}
  \caption{Top panel: Sum of PAH luminosities versus the H$_2$
    emission. Bottom panel:  the 11.3$\mu$m PAH equivalent width with
      respect to the hardness of the radiation field as given by the
      \ion{[Ne}{3]}15.5$\mu$m/\ion{[Ne}{2]}12.8$\mu$m ratio.}
\end{figure}

\begin{figure*}
  \begin{center}
  \includegraphics[width=14cm]{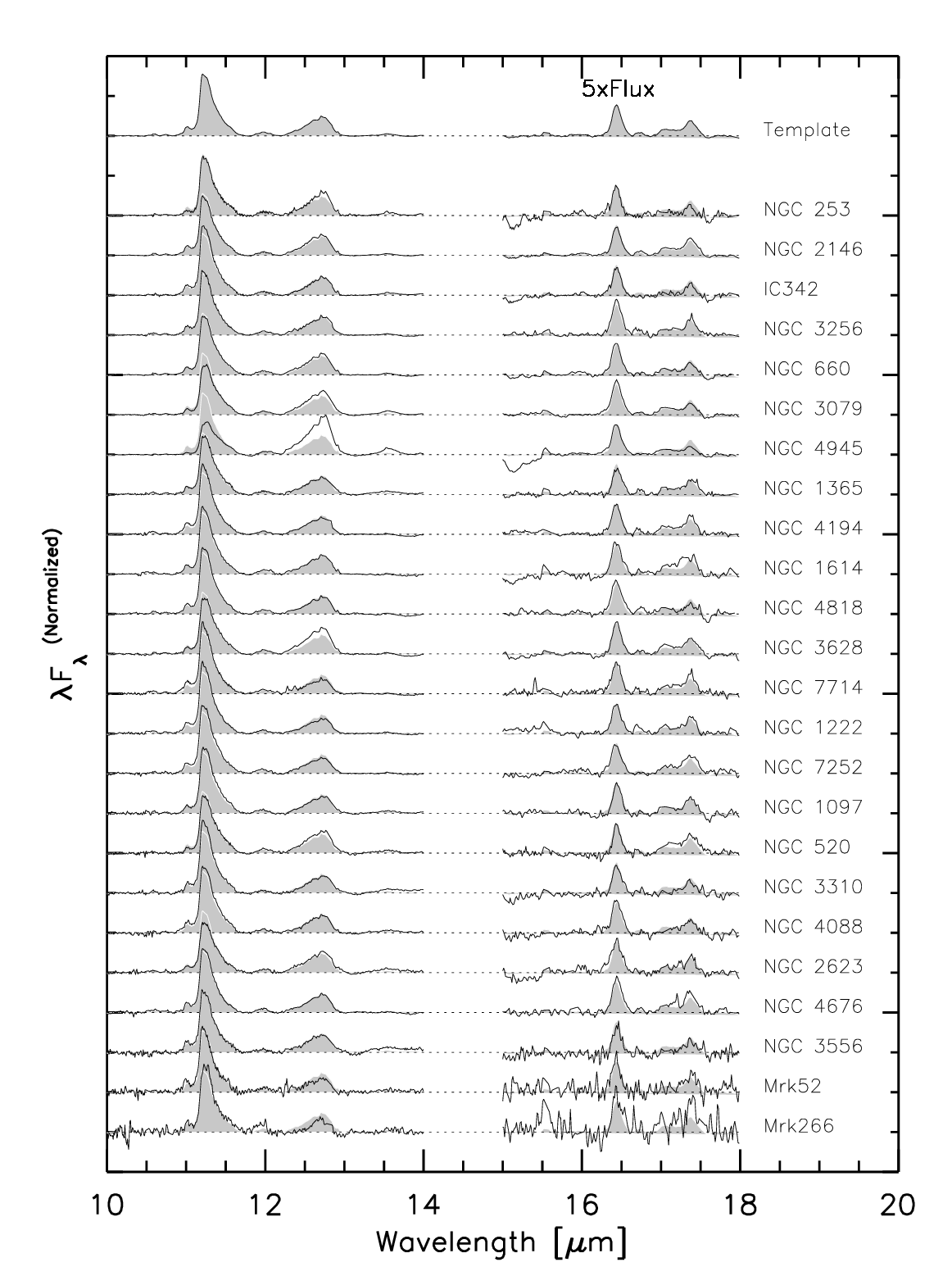}
  \end{center}
  \caption{Continuum subtracted spectra of the starburst sample. The
    thin black line represents the spectrum of the corresponding
    object and the grey shaded area that of the template (with a white
    edge to distinguish the grey areas in the overlapping 11.3$\mu$m
    band). The fine-structure line emission has been removed to better
    show the PAH profile in the starburst spectra. The spectra have
    been normalized to the total PAH emission and ordered according to
    their 11.3~$\mu$m PAH strength (strongest at bottom), and the
    15-18$\mu$m flux density has been multiplied by a factor 5 to
    better display the PAHs in that region.}
\end{figure*}

\begin{deluxetable}{l c c c}
\tabletypesize{\scriptsize}
  \tablewidth{0pt}
  \tablecaption{H$_2$ conditions.\label{lines_h2_t}}
  \tablehead{\colhead{Object} & \colhead{T$_{ex}$\tablenotemark{a}}  &
    \colhead{N(H$_2$)}  &    \colhead{H$_2$ Mass}\\
    &   \colhead{(K)}  &  \colhead{$\times$10$^{19}$ cm$^{-2}$} & \colhead{$\times$10$^{5}$M$_{\odot}$}}
  \startdata

     NGC\,253  &    523.7  &    49.9  &     0.2 \\
     NGC\,520  &    282.5  &    22.7  &    10.1 \\
     NGC\,660  &    358.9  &    29.1  &     2.2 \\
    NGC\,1097  &    340.6  &    18.1  &     2.5 \\
    NGC\,1222  &    277.1  &    21.0  &    10.8 \\
    NGC\,1365  &    342.8  &    24.9  &     3.9 \\
      IC\,342  &    492.5  &     9.2  &     0.1 \\
    NGC\,1614  &    363.4  &    11.4  &    21.9 \\
    NGC\,2146  &    343.2  &    35.2  &     4.7 \\
    NGC\,2623  &    322.7  &    11.1  &    32.7 \\
    NGC\,3079  &    378.2  &    45.9  &     7.0 \\
    NGC\,3256  &    360.5  &    34.4  &    21.2 \\
    NGC\,3556  &    298.6  &     5.3  &     0.5 \\
    NGC\,3628  &    311.6  &    26.8  &     1.3 \\
    NGC\,4088  &    372.6  &     5.5  &     0.5 \\
    NGC\,4194  &    362.8  &    11.2  &     9.0 \\
      Mrk\,52  &    472.9  &     1.5  &     0.7 \\
    NGC\,4676  &    307.6  &     8.8  &    38.3 \\
    NGC\,4818  &    292.2  &    32.4  &     1.4 \\
    NGC\,4945  &    393.1  &    57.4  &     0.4 \\
     Mrk\,266  &    326.6  &    12.5  &    82.4 \\
    NGC\,7252  &    357.7  &     6.4  &    13.8 \\
    NGC\,7714  &    399.5  &     4.6  &     3.3 \\

  \enddata
\tablenotetext{a}{Based on the 0-0~S(1) and S(2) transitions.}
\end{deluxetable}

    Figure~12 displays the relation between the PAH features and the
    emission lines. In the top panel, the sum of the PAH luminosities
    is plotted against the sum of the H$_2$ lines. Both are produced
    in the PDR, albeit in different regions, and their fluxes
    correlate well.  In the bottom panel the equivalent width (EW) of
    the 11.3~$\mu$m PAH\footnote{For simplicity we use the equivalent
      width of the combined 11.0 and 11.3~$\mu$m components. This is
      valid as both bands correlate well with each other (Figure~11).}
    is plotted against the \ion{[Ne}{3]} over \ion{[Ne}{2]} line
    ratio.  There is no correlation, indicating that there is no
    dependence of the PAH with the hardness of the radiation field. In
    fact, the EW of the 11.3$\mu$m shows little variation in
    Figure~12.  In AGNs, the PAH EW is seen to decrease with harder
    radiation field as the PAH are further ionized or destroyed, but
    the PAH EW is flat for soft radiation fields as those in
    Starbursts (Lebouteiller et al., in prep.).

    Another characteristic of the PAHs is their profile, which has
    been found to change between and within sources. This led
    \citet{pee02} to establish a classification based on the peak
    position of the different bands. Figure~13 shows the continuum
    subtracted spectra for all the objects in the sample in the
    regions of interest (where the emission lines have been removed).
    The grey area in the figure represents the template spectrum.  As
    opposed to the template shown in Figures~3 and 4, this template
    has been normalized to the sum of the PAHs (and not the sum of the
    PAHs plus lines), but the are both very similar. This PAH
    continuum subtracted spectrum of the template starburst is also
    available in electronic format (tab26.txt).
  The fact that the spectrum of each object (thin black line in the
  figure) follows that of the template (grey area) indicates there is
  hardly any variations in the profiles.

  The relative strengths of the features is also the same for most of
  the object although there are a few variations.  The most striking
  is 12.7$\mu$m PAH in NGC\,4945 which is stronger than the 11.3$\mu$m
  PAH. The reason for this is the stronger silicate dust extinction on
  the 11.3$\mu$m PAH than the 12.7$\mu$m PAH. This band is also strong
  (relative to the 11.3$\mu$m PAH) in those starburst with AGN
  signature (e.g.  NGC\,660, NGC\,3079, NGC\,3628) but also in a few
  objects with no AGN (e.g.  NGC\,253).
  The 17.0$\mu$m complex does not always follow that trend.  Despite
  these differences the profiles are near akin. As we saw from
  previous sections the metallicities are similar (within a few factor
  of each other), and the radiation fields are low in most of them.
  Therefore, it is possible that the invariability in the PAH spectra
  of the starburts in our sample are probably due to the similar
  conditions in these galaxies.

  \subsection{Molecular Hydrogen}

  We detect three pure-rotational transitions 0-0~S(2), S(1),
  and S(0) of H$_2$ at 12.28, 17.03, and 28.21~$\mu$m respectively.
  The S(1) transition at 17.03~$\mu$m is the strongest of the three.
  Figure~14 shows that, while there is some scatter, there is a good
  correlation of the H$_2$ lines with the \ion{[Si}{2]} line. This is
  expected since both originate in the PDR. In fact, \ion{[Si}{2]} is
  an important fine-structure cooling line of the PDR. We note that in
  order to combine the three lines we used a scaling factor between
  the SH and LH modules determined by matching the continuum in the
  overlap region in both modules around 19.5~$\mu$m. This is not ideal
  since no off observations were subtracted and it contributes
  to the scatter present in the figure. The ratio of the sum of the
  H$_2$ lines over the \ion{[Si}{2]} line is 0.22, with values ranging
  from 0.04 to 0.5.  This is very similar to the values found by
  \citet{Roussel07} in their starburst (SINGS) sample.

  The molecular hydrogen lines can be used to give information on the
  conditions of the warm component in the PDR. There is a relation
  between the density and temperature assuming local thermal
  equilibrium and using the Boltzman equation \citep{anc00,ber05}.
  Using this relation and the lines in the SH module (to avoid scaling
  effects), we have derived the excitation temperature (T$_{ex}$),
  column density N(H$_{2}$), and warm molecular mass for the sample.
  These are given in Table~6. The average excitation temperature and
  column density are 360$\pm$64~K and 2$\times$10$^{20}$~cm$^{-2}$
  respectively, and the warm molecular mass ranges from several 10$^4$
  to 8$\times$10$^6$M$_{\odot}$, with an average mass of
  1.1$\times$10$^{6}$M$_{\odot}$. We stress that these numbers
  correspond only to the region covered by the SH slit and are based
  on the fluxes for just the S(1) and S(2) lines.

\begin{figure}
  \begin{center}
  \includegraphics[width=6.5cm,angle=90]{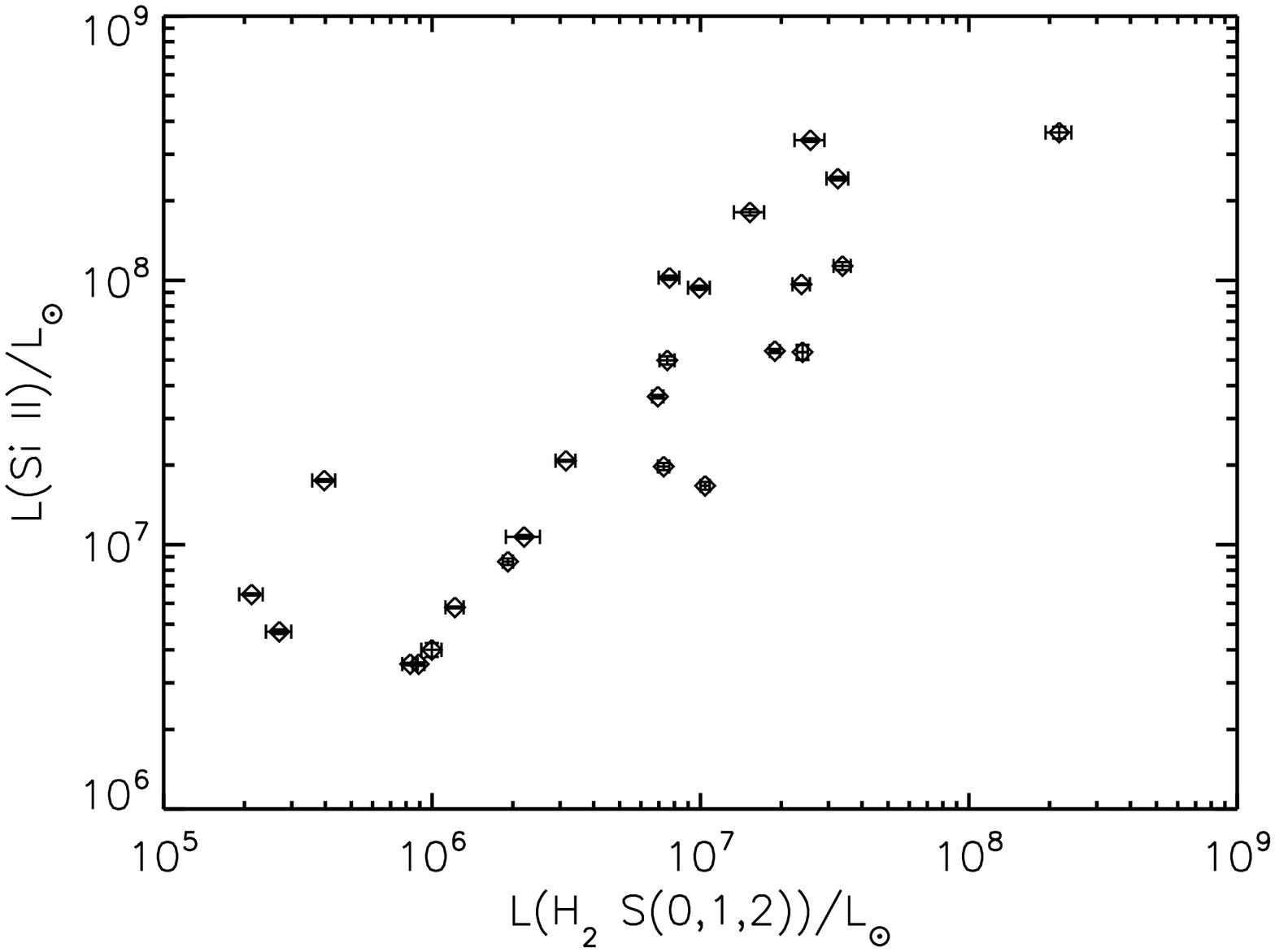}
  \end{center}
  \caption{Luminosities of the \ion{[Si}{2]} line versus the sum of
    the three H$_2$ rotational lines measured in the IRS spectrum. For
    this plot the fluxes of the H$_2$ S(2) and S(1) lines have been
    scaled based on the SH to LH continuum (see text for further
    details).}
\end{figure}

  Recently, \citet{Roussel07} studied the warm molecular hydrogen
  emission in the nuclear region of a sample of SINGS galaxies.  The
  range in areas covered in their observations (60~pc to 3.8~kpc) is
  very similar to the range covered in our study (see \S2). Their
  column densities compare well to those found for our starburst
  galaxies, but their excitation temperatures are on average cooler by
  $\sim$150~K, and their masses significantly larger
  (10$^{5}$-10$^{8}$ M$_{\odot}$).  This is expected because the SINGS
  galaxies have a much lower star formation rate per unit volume
  compared to our starbursts, and since they are less luminous, a
  larger fraction of their molecular Hydrogen is colder. `Cold'
  molecular hydrogen is traced by the S(0) line which is not used in
  our analysis as it falls in the LH slit, but it is used in the SINGS
  sample where aperture size differences do not exist.  Hence, our
  computed $H_{2}$ masses are lower than those found for the SINGS
  sample, but this is likely an artifact of the absence of the S(0)
  line in our calculations, rather than a real difference.

  We find no relation between the excitation temperature and sources
  with and without AGNs.  \citet{Higdon06} analyzed {\em Spitzer} data
  of a sample of ULIRGs. Their average excitation temperature
  (336$\pm$15~K) agrees with that of our starburst sample, but again
  their warm molecular masses are several hundred's times larger
  (2$\times$10$^8$M$_{\odot}$).  It is important to note that most
  ULIRGs fit in the SH aperture while in the starbursts we are only
  measuring the central region. Still, this difference may point as
  well to a larger fraction of mass being colder in the Starburst
  sample.



\section{Conclusions}\label{conclusion}

\begin{enumerate}

\item We have presented high resolution spectra (R$\sim$600) of a
  sample (24) of nearby starburst galaxies.  The spectra are dominated
  by emission from forbidden fine-structure lines and polycyclic
  aromatic hydrocarbons features.

\item \ion{[Ne}{5]} is detected in six of the eight starbursts with a
  previously known AGN component. This line is also seen in NGC\,4194.
  
\item Several PAH features have been measured in the spectra, some of
  which are weak and are rarely seen in the spectra of starburst
  galaxies and HII regions. 

\item An emission feature around 10.7$\mu$m and an absorption feature
  at 13.7$\mu$m are detected for the first time.  Their identification
  is unknown, but the strength of the feature at 10.7$\mu$m correlates
  with that other PAH features.

\item The average conditions of the warm gas as traced by the S(1) and
  S(2) pure rotational molecular hydrogen lines indicates an average
  excitation temperature of 355$\pm$60~K and a density of
  2$\times$10$^{20}$~cm$^{-2}$. The warm molecular mass ranges from
  several 10$^4$ to 4$\times$10$^6$~M$_{\odot}$.

\item We provide the combined average spectrum of the 15 `pure'
  starburst galaxies of our sample, which is likely the highest signal
  to noise mid-infrared spectrum of a $\sim 10^{11}$ L$_{\odot}$
  starburst galaxy available from Spitzer. This template is
  representative of a large range in luminosity and IR color, and
  provides a good indication of the spectral properties of starburst
  galaxies at high redshift.


\end{enumerate}





\acknowledgements VC acknowledges partial support from the EU grants
ToK 39965 and FP7-REGPOT 206469. We thank Ian Waters and Hongyu Xiao
for their help with the line measurement.



\clearpage


\begin{thebibliography}{}

\bibitem[Allamandola et al.(1985)]{all85} Allamandola, L.J., Tielens,
  A.G.G.M., Barker, J.R. 1985, \apjl, 290, L25

\bibitem[Armus et al.(2007)]{Armus07} Armus, L., Charmandaris, V.,
  Bernard-Salas, J., et al. 2007, \apj, 656, 148

\bibitem[Asplund et al.(2005)]{asp05} Asplund, M., Grevesse, N., \&
  Sauval, A.J. 2005, ASPC 336, 25

\bibitem[Asplund et al.(2006)]{asp06} Asplund, M., Grevesse, N., \&
  Sauval, A.J. 2006, Commun. Asteroseismology, 147, 76

\bibitem[Bernard-Salas et al.(2001)]{ber01} Bernard-Salas, J.,
  Pottasch, S.R., Beintema, D.A., \& Wesselius, P.R. 2001, \aap, 367, 949 

\bibitem[Bernard-Salas et al.(2008)]{ber08} Bernard-Salas, J.,
  Pottasch, S.R., Gutenkunst, S., Morris, P.W., \& Houck, J.R. 2008, \apj,
  672, 274

\bibitem[Bernard-Salas \& Tielens(2005)]{ber05} Bernard-Salas, J., \&
  Tielens, A.G.G.M. 2005, \aap, 431, 523


\bibitem[Blain et al.(2002)]{Blain02} Blain, A.~W., Smail, I., Ivison,
  R.J., Kneib, J.-P., \& Frayer, D.T. 2002, \physrep, 369, 111

\bibitem[Boulanger et al.(1998)]{bou98} Boulanger, F., Boissel, P.,
  Cesarsky, D., \& Ryter, C. 1998, \aap, 339, 194

\bibitem[Brandl et al.(2006)]{Brandl06} Brandl, B.R., Bernard-Salas,
  J., Spoon, H.W.W., et al. 2006, \apj, 653, 1129

\bibitem[Cohen et al.(2003)]{coh03} Cohen, M., Megeath, T.G.,
  Hammersley, P. L., Martin-Luis, F., \& Stauffer, J. 2003, \aj, 125,
  2645

\bibitem[Dale et al.(2006)]{Dale06} Dale, D.A., Smith, J.D.T., Armus,
  L., et al. 2006, \apj, 646, 161

\bibitem[Dale et al.(2009)]{dal09} Dale, D.A., Smith, J.D.T.,
  Schlawin, E.A., et al. 2009, \apj, 693, 1821


\bibitem[Dudik et al. (2007)]{Dudik07} Dudik, R.P., Weingartner, J.C.,
  Satyapal, et al. 2007, \apj, 664, 71

\bibitem[Elbaz \& Cesarsky(2003)]{Elbaz03} Elbaz, D., \& Cesarsky,
  C.J. 2003, Science, 300, 270


\bibitem[Farrah et al.(2007)]{Farrah07} Farrah, D., Bernard-Salas, J.,
  Spoon, H.W.W., et al. 2007, \apj, 667, 149

\bibitem[Farrah et al.(2008)]{Farrah08} 
Farrah, D., et al.\ 2008, \apj, 677, 957 

\bibitem[Feldman \& Widing(2003)]{fel03} Feldman, U., \& Widing, K.G.
  2003, Sp.Sci.Rev. 107, 665

\bibitem[F{\"o}rster Schreiber et al.(2004)]{Forster04} F{\"o}rster
  Schreiber, N.~M., Roussel, H., Sauvage, M., \& Charmandaris, V.
  2004, \aap, 419, 501

\bibitem[Galliano et al.(2008)]{gal08} Galliano, F., Madden, S.C.,
  Tielens, A.G.G.M., Peeters, E., \& Jones, A.P. 2008, \apj, 679, 310

\bibitem[Genzel et al.(1998)]{genzeletal98} Genzel, R., Lutz, D.,
  Sturm, E., et al.  1998, \apj, 498, 579

\bibitem[Gonzalez-Delgado et al.(1995)]{gon95} Gonzalez-Delgado, R.M.,
  Perez, E., Diaz, A.I., et al. 1995, \apj, 439, 604

\bibitem[Grevesse \& Sauval(1998)]{gre98} Grevesse, N., \& Sauval, A.J.
  1998, Space Sci. Rev., 85, 161

\bibitem[Groves et al.(2008)]{gro08} Groves, B., Nefs, B., \& Brandl,
  B.R. 2008, \mnras, 391, L113 

\bibitem[Hao et al. (2009)]{hao09} Hao, L., Wu, Yanling, Charmandaris,
  V., et al. (2009), \apj, submitted

\bibitem[Henry \& Kwitter(2004)]{hen04} Henry, R.B.C., Kwitter, K.B.,
  \& Balick, B. 2004, AJ 127, 2284

\bibitem[Higdon et al.(2006)]{Higdon06} Higdon, S.J.U., Armus, L.,
  Higdon, J.L., Soifer, B.T., \& Spoon, H.W.W. 2006, \apj, 648, 323

\bibitem[Higdon et al.(2004)]{smart} Higdon, S.J.U., Weedman, D.,
  Higdon, J.L., et al. 2004, \pasp, 116, 975

\bibitem[Ho \& Keto(2007)]{ho07} Ho, L.C., \& Keto, E. 2007, \apj,
  658, 314 

\bibitem[Hony et al.(2001)]{hon01} Hony, S., van kerckhoven, C.,
  Peeters, et al. 2001, \aap, 370, 1030

\bibitem[Houck et al.(2004)]{Houck04} Houck, J.R., Roellig, T.L., Van
  Cleve, J., et al. 2004, \apjs, 154, 18

\bibitem[Hunter \& Kaufman(2007)]{hun07} Hunter, D.A., \& Kaufman,
  M. 2007, \aj, 134, 721

\bibitem[Kennicutt(1998)]{ken98} 
Kennicutt, R.~C., Jr.\ 1998, \araa, 36, 189 

\bibitem[Kessler et al.(1996)]{kess96} 
Kessler, M.~F., et al.\ 1996, \aap, 315, L27 

\bibitem[Lagache et al.(2005)]{lag05} 
Lagache, G., Puget, J.-L., \& Dole, H.\ 2005, \araa, 43, 727 

\bibitem[Laurent et al.(2000)]{Laurent00} Laurent, O., Mirabel, I.F.,
  Charmandaris, V., et al. 2000, \aap, 359, 887

\bibitem[Le Floc'h et al.(2005)]{lef05} Le Floc'h, E., Papovich, C.,
  Dole, H., et al. 2005, \apj, 632, 169

\bibitem[Lebouteiller et al.(2008)]{leb08} Lebouteiller, V.,
  Bernard-Salas, J., Brandl, B., et al. 2008, \apj, 680, 398

\bibitem[Leger \& Puget(1984)]{Leger84} Leger, A., \& Puget, J.L.\
  1984, \aap, 137, L5

\bibitem[Li \& Draine(2001)]{li01} Li, A., \& Draine, B.T. 2001, \apj,
  554, 778


\bibitem[Lonsdale et al.(2006)]{lfs06}
Lonsdale, C.~J., Farrah, D., \& Smith, H.~E.\ 2006, Astrophysics Update 2, 285

\bibitem[Lutz et al.(2003)]{lutzetal03} Lutz, D., Sturm, E., Genzel,
  R., et al. 2003, \aap, 409, 867

\bibitem[Marigo et al.(2003)]{mar03} Marigo, P., Bernard-Salas, J.,
  Pottasch, S.R., Tielens, A.G.G.M., \& Wesselius, P.R., 2003, \aap,
  409, 619

\bibitem[Mart\'{i}n-Hern\'andez et al.(2002)]{mar02}
  Mart\'{i}n-Hern\'andez, N.L., Peeters, E., Morisset, C. et al. 2002,
  \aap, 381, 606

\bibitem[Neugebauer et al.(1984)]{neu04} Neugebauer, G., et 
al.\ 1984, \apjl, 278, L1 

\bibitem[Oliva et al.(1999)]{oli99} Oliva, E., Moorwood, A.F.M.,
  Drapatz, S., Lutz, D., \& Sturm, E. 1999, \aap, 343, 943

\bibitem[Pastoriza et al.(1993)]{pas93} Pastoriza, M.G., Dottori,
  H.A., Terlevich, E., Terlevich, R., \& Diaz, A.I. 1993, \mnras, 260,
  177

\bibitem[Peeters et al.(2002)]{pee02} Peeters, E., Hony, S., Van
  Kerckhoven, C., et al. 2002, \aap, 390, 1089

\bibitem[Peeters et al.(2004)]{Peeters04} Peeters, E., Spoon, H.W.W.,
  \& Tielens, A.G.G.M. 2004, \apj, 613, 986


\bibitem[Pilyugin et al.(2004)]{pil04} Pilyugin, L.S., Contini, T., \&
  V\'{i}lchez, J.M. 2004, \aap, 423, 427

\bibitem[Pottasch \& Bernard-Salas(2006)]{pot06} Pottasch, S.R., \&
  Bernard-Salas, J. 2006, \aap, 457, 189

\bibitem[Rigopoulou et al.(1999)]{rigopoulouetal99} Rigopoulou, D.,
  Spoon, H.W.W., Genzel, R., et al. 1999, \aj, 118, 2625


\bibitem[Rosenberg et al.(2008)]{Rosenberg08} Rosenberg, J.L., Wu, Y.,
  Le Floc'h, E., et al. 2008, \apj, 674, 814

\bibitem[Roussel et al.(2007)]{Roussel07} Roussel, H., Helou, G.,
  Hollenbach, D.J., et al.\ 2007, \apj, 669, 959

\bibitem[Roussel et al.(2001)]{Roussel01} Roussel, H., Sauvage, M.,
  Vigroux, L., \& Bosma, A. 2001, \aap, 372, 427

\bibitem[Rubin et al.(2008)]{rub08} Rubin, R.H., Simpson, J.P.,
  Colgan, S.W.J., et al. 2008, \mnras, 387, 45 

\bibitem[Rubin et al.(2007)]{rub07} Rubin, R.H., Simpson, J.P.,
  Colgan, S.W.J., et al. 2007, \mnras, 377, 1407 

\bibitem[Sloan et al.(1999)]{slo99} Sloan, G.C., Hayward, T.L.,
  Allamandola, L.J., et al. 1999, \apj, 513, L65

\bibitem[Smith et al.(2004)]{smi04} Smith, J.D.T., Dale, D.A., Armus,
  L., et al. 2004, \apjs, 154, 199

\bibitem[Smith et al.(2007)]{Smith07} Smith, J.D.T., Draine, B.T.,
  Dale, D.A., et al. 2007, \apj, 656, 770

\bibitem[Spoon et al.(2007)]{Spoon07} Spoon, H.W.W., Marshall, J.A.,
  Houck, J.R., et al. 2007, \apjl, 654, L49

\bibitem[Spoon et al.(2009, in prep)]{Spoon09} Spoon, H.W.W., et al. 2009,\apj
  in prep.

\bibitem[Stasinska (1990)]{sta90} Stasinska, G., 1990, \aaps, 83, 501

\bibitem[Sturm et al.(2002)]{sturmetal02} Sturm, E., Lutz, D., Verma,
  A., et al. 2002, \aap, 393, 821

\bibitem[Thornley et al.(2000)]{thornleyetal00} Thornley, M.D.,
  Schreiber, N.M.F., Lutz, D., et al. 2000, \apj, 539, 641

\bibitem[van den Ancker et al.(2000)]{anc00} van den Ancker, M.E.,
  Tielens, A.G.G.M., \& Wesselius, P.R. 2000, \aap, 358, 1035 

\bibitem[Van Kerckhoven et al.(2000)]{ker00} Van Kerckhoven, C., Hony,
  S., Peeters, E., et al. 2000, \aap, 357, 1013

\bibitem[Verma et al.(2003)]{vermaetal03} Verma, A., Lutz, D., Sturm,
  E., et al. 2003, \aap, 403, 829

\bibitem[Vermeij \& van der Hulst(2002)]{ver02} Vermeij, R., \& van
  der Hulst, J.M. 2002, \aap, 391, 1081

\bibitem[Wang \& Liu(2008)]{wan08} Wang, W., \& Liu, X.-W. 2008,
  /mnras, 389, 33

\bibitem[Weedman et al.(1981)]{Weedman81} Weedman, D. W., Feldman, F.
  R., Balzano, V. A., et al., 1981, \apj, 248, 105

\bibitem[Weedman et al.(2005)]{Weedman05} Weedman, D.W., Hao, L.,
  Higdon, S.J.U., et al. 2005, \apj, 633, 706

\bibitem[Werner et al.(2004)]{wer04} Werner, M., Roellig, T. L., Low,
  F.  J., et al. 2004, \apjs, 154, 1

\bibitem[Wu et al.(2008)]{wu08} Wu, Y., Bernard-Salas, J.,
  Charmandaris, V., et al. 2008, \apj, 673, 193

\bibitem[Wu et al.(2006)]{Wu06} Wu, Y., Charmandaris, V., Hao, L.,
  Brandl, B.R., Bernard-Salas, J., Spoon, H.~W.~W., \& Houck, J.R.
  2006, \apj, 639, 157

\end{thebibliography}
\end{document}